\newif\ifsingle

\ifsingle
\documentclass[11pt,draftclsnofoot, onecolumn]{IEEEtran}		
\else		
\documentclass[10pt,final, twocolumn]{IEEEtran}
\fi

\usepackage{Latent}

\IEEEoverridecommandlockouts 
 
\title{Latent-KalmanNet: Learned Kalman Filtering for Tracking from High-Dimensional Signals}

\author{  
	\IEEEauthorblockN{Itay Buchnik,  \IEEEmembership{Student Member, IEEE}, Damiano Steger,  \IEEEmembership{Student Member, IEEE}, \\ Guy Revach,  \IEEEmembership{Student Member, IEEE}, Ruud J. G. van Sloun,  \IEEEmembership{Member, IEEE}, \\ Tirza Routtenberg,  \IEEEmembership{Senior Member, IEEE}, and Nir Shlezinger,  \IEEEmembership{Member, IEEE}}
\thanks{
Parts of this work were accepted for presentation at the IEEE International Conference on Acoustics, Speech, and Signal Processing (ICASSP) 2023 as the paper~\cite{LatentICASSP}.
 I. Buchnik, T. Routtenberg, and  N. Shlezinger are with the School of ECE, Ben-Gurion University of the Negev, Be'er Sheva, Israel (e-mail: itaybuch@post.bgu.ac.il; \{tirzar; nirshl\}@bgu.ac.il). T. Routtenberg is also with the ECE Department, Princeton University, Princeton, NJ. 
 D. Steger and G. Revach are with the D-ITET, ETH Zürich, Switzerland (e-mail: stegerd@student.ethz.ch; grevach@ethz.ch). R. J. G. van Sloun is with the EE Dpt., Eindhoven University of Technology, The Netherlands (e-mail: r.j.g.v.sloun@tue.nl). This work is partially supported by the Israeli Ministry of National Infrastructure, Energy, and Water Resources. We thank Hans-Andrea Loeliger for the helpful discussions.
}}

\begin{document}

\maketitle
\begin{abstract} 

The \ac{kf} is a widely-used algorithm for tracking dynamic systems that are captured by \ac{ss} models. The need to fully describe a \ac{ss} model limits its applicability under complex settings, e.g., when tracking based on visual data, and the processing of high-dimensional signals often induces notable latency. These challenges can be treated by mapping the measurements into latent features obeying some postulated closed-form \ac{ss} model, and applying the \ac{kf} in the latent space. However, the validity of this approximated \ac{ss} model may constitute a limiting factor. In this work, we study tracking from high-dimensional measurements under complex settings using a hybrid model-based/data-driven approach. By gradually tackling the challenges in handling the observations model and the  task, we develop \name, which implements tracking from high-dimensional measurements by leveraging data to jointly learn the \ac{kf}  along with the latent space mapping. \name~combines a learned encoder with data-driven tracking in the latent space using the recently proposed-\acl{kn}, while identifying the ability of each of these trainable modules to assist its counterpart via providing a suitable prior (by \acl{kn}) and by learning a latent representation that facilitates data-aided tracking (by the encoder). 
Our empirical results demonstrate that the proposed \name~achieves improved accuracy and run-time performance over both model-based and \acl{dd} techniques by learning a surrogate latent representation that most facilitates tracking, while operating with limited  complexity and  latency. 
\end{abstract}
 
\acresetall

\section{Introduction}
\label{sec:intro}
Tracking the hidden state of dynamic systems is a fundamental problem in various fields, including signal processing, control, and finance. In many real-world applications, such as  autonomous driving, smart city monitoring, and visual surveillance, tracking is based on noisy high-dimensional observations, e.g., visual data. The classic \ac{kf} \cite{kalman1960new} algorithm and its variants \cite[Ch. 10]{durbin2012time} have been the go-to approach for tracking, relying on the representation of the dynamics as a \ac{ss} model that describes the state evolution and the sensing model. \ac{kf} is widely used due to its computational efficiency and optimality properties. However, the reliance of the \ac{kf} and its variants on an accurate description of the underlying dynamics as a closed-form \ac{ss} model with Gaussian noise restricts it applicability when tracking from complex high-dimensional data.


In particular, the \ac{kf}  assumes linear dynamics with Gaussian noise of a known distribution. Variations of the \ac{kf}, such as the \ac{ekf} \cite{ larson1967application}  and the unscented Kalman filter \cite{wan2001unscented}, can cope with nonlinear Gaussian \ac{ss} models, yet they require an accurate description of the nonlinearities, which is often unavailable when  dealing with visual data, and their complexity grows when processing high-dimensional observations. Alternative tracking methods based on Bayesian filtering  \cite{lin2019enllvm,  djuric2003particle, gordon1993novel} do not assume Gaussian modeling, yet are often computationally complex.
While for certain families of high-dimensional observations, such as graph signals, one can leverage structures in the data to notably reduce tracking complexity \cite{isufi2020observing, sagi2022gsp,GSPICASSP}, such approaches  do not naturally extend to  other domains of high-dimensional data. 
Moreover, all of the aforementioned techniques are model-based, relying on full knowledge of the \ac{ss} model, which is likely to be unavailable when tracking based on high-dimensional measurements such as visual data.

In recent years, the combination of large-scale datasets and advancements in deep learning has led to the development of several \acl{dd} filtering methods, see review in \cite{cheng2023machine}. These methods have shown empirical success in processing visual data, and typically involve \ac{dnn} architectures, such as \acp{rnn}  \cite{gu2017dynamic}, attention mechanisms \cite{tang2021probabilistic}, and deep Markov models \cite{zhi2020factorized} for state tracking tasks. While these methods are based on architectures designed for generic time sequence processing, several \ac{dnn} architectures were proposed specifically for tracking in \ac{ss} models being inspired by model-based tracking algorithms~\cite{rangapuram2018deep, millidge2021neural, jouaber2021nnakf, ruhe2021self, Ruhe2021SelfSupervisedII,becker2019recurrent, nguyen2021switching,klushyn2021latent, wang2018visual}, resulting in, e.g., \acp{dnn} whose internal interconnection follows the flow of the \ac{ekf}. Among these existing works, the systems of \cite{becker2019recurrent, nguyen2021switching,klushyn2021latent, wang2018visual} were specifically designed to cope with high-dimensional measurements, with a leading basis architecture being the \ac{rkn} of \cite{becker2019recurrent}. However, those methods suffer from difficulty in training, sensitivity to initialization and generalization problems. Moreover, were not leverage domain knowledge regarding the state evolution, even when such is available, as is the case in various applications including, e.g., localization and navigation~\cite[Ch. 6]{bar2004estimation}. 

While the above \acl{dd} approaches do not use knowledge about the \ac{ss} model, one can combine model-agnostic deep-learning tools with \ac{ss}-aware processing. A candidate approach to do so in the context of high dimensional data is to use a \ac{dnn} decoder to capture the complex observation model~\cite{zhou2020kfnet}, thus overcoming the need to analytically describe it, yet preserving the complexity associated with tracking using high-dimensional data. 
Alternatively, a widely adopted approach encodes the observations into a latent space via a \ac{dnn}, i.e., using instead the inverse of the observations model.  These latent features are then used to track the state with, e.g., a conventional \ac{kf}. This approach assumes that the latent features obey a simple \ac{ss} model, typically a known Gaussian one in the latent domain as in \cite{zhou2020kfnet, coskun2017long, fraccaro2017disentangled, Li2021ReplayOL, krishnan2015deep, laufer2018hybrid}. However, the resulting latent \ac{ss} model is often non-Gaussian, which can impact the tracking accuracy in the latent space. 

In this work, we propose \name, which addresses the difficulties of tracking high-dimensional data by simultaneously learning to track along with the latent space representation. To achieve this, we utilize the recently proposed \acl{kn} \cite{revach2022kalmannet,klein2022uncertainty,revach2022unsupervised}, that learns from data to perform Kalman filtering in partially known \ac{ss} models as a form of model-based deep learning  \cite{shlezinger2020model, shlezinger2022model}. \acl{kn} relies on a (possibly approximated) description of the sensing function. However, this information is unavailable in the setting considered in this work of complicated high-dimensional data, and the solution complexity grows with the dimensions of the measurements. Therefore, in \name~we combine \acl{kn} with latent-space encoding, and propose a novel training method which {\em jointly learns the latent representation and filter operation}. \name~uses a latent transformation is assisted by its subsequent data-aided tracking method. The resulting latent representation is suitable for tracking while maintaining the interpretability and low complexity of the \ac{kf}. 

In particular, we first identify the main challenges associated with tracking from high-dimensional measurements in $(1)$ the need to model stochasticity; $(2)$ the operation with a possibly intractable measurements model; $(3)$ the need to be applicable in real-time; and $(4)$ the presence of possible mismatches in the state evolution model. Based on these challenges, we derive \name~by gradually addressing each specific challenge, while accounting for settings where the state can be either partially observable or fully observable. The resulting \name~combines two trainable components -- an encoder that maps the observations into a latent representation, and \acl{kn}, which tracks based on the latent features. Instead of designing these components separately, we exploit the ability of each module to facilitate the operation of its counterpart. Specifically, the tracking module is used to provide a prior for encoding, while the encoder generates a latent representation that is most suitable for tracking, and this desired behavior is learned from data using a dedicated alternating training mechanism.  
Our experimental study evaluates \name~for tracking in challenging settings with high-dimensional visual data, identifying the  benefits of each of the components incorporated in \name, while showing that the synergistic design of latent encoding and tracking yields notable performance improvements. Furthermore, we demonstrate that \name~outperforms  classic model-based nonlinear tracking algorithms as well as state-of-the-art deep architectures in terms of state estimation accuracy as well as inference speed.

The rest of the paper is organized as follows: Section \ref{sec:System_Model_and_Preliminaries} details the problem formulation and briefly describes the \ac{ekf} and \acl{kn}. Section \ref{sec:Learned_Kalman_Filtering_in_the_Latent_Space} presents  \name~in a step-by-step manner, along with the proposed training method. Our numerical evaluation is provided in Section \ref{sec:Empirical_Study}, while Section~\ref{sec:conclusions} concludes the paper.

Throughout the paper, we use boldface lower-case letters for vectors and boldface upper-case letters for matrices. The transpose, $\ell_2$ norm, and gradient operator are denoted by $\set{\cdot}^\top$,  $\norm{\cdot}$, and  $\nabla_{(\cdot)}$, respectively. 
Finally, $\greal$ and $\gint$ are the sets of real and integer numbers, respectively.

\section{System Model and Preliminaries}\label{sec:System_Model_and_Preliminaries}
In this section, we present the system model and relevant preliminaries needed to derive \name~in Section~\ref{sec:Learned_Kalman_Filtering_in_the_Latent_Space}. We start by formulating the problem of tracking in partially known \ac{ss} models with high-dimensional observations in Subsection~\ref{ssec:Problem_Formulation}. Then, in Subsection~\ref{ssec:Data-Driven_Filtering_with_KalmanNet}, we briefly recall the model-based \ac{ekf} and \acl{kn}  of \cite{revach2022kalmannet}, and identify their shortcomings for the considered setting.

\subsection{Problem Formulation}\label{ssec:Problem_Formulation}
We consider a dynamic system characterized by a (possibly) nonlinear, continuous \ac{ss} model in discrete-time  $t\in \mathbb{Z}$. 
Let $\gvec{x}_t$ be the $m\times 1$ state vector, which evolves by a nonlinear state evolution function $\gvec{f}(\cdot)$, and is driven by an additive zero-mean noise $\gvec{e}_t$. The $n\times 1$ observation vectors $\gvec{y}_t$, $t\in \mathbb{Z}$, are high-dimensional, and in particular $n\gg m$, that can be, e.g., the vector representation of an image/tensor\footnote{{For mathematical simplicity, we formulate our high-dimensional observations in vector form, which also represents tensor data by stacking their elements in vector form. The size $n$ considered is the total number of elements in the observation.}}.  The observed $\gvec{y}_t$ is related to the state $\gvec{x}_t$ via a complex and possibly unknown measurement function $\gvec{h}(\cdot)$ with additive zero-mean noise $\gvec{v}_t$. The resulting \ac{ss} model is given by:
\begin{subequations}
\label{eqn:SSModel}
\begin{align}
\gvec{x}_{t}&= \gvec{f}\brackets{\gvec{x}_{t-1}}+\gvec{e}_t, 
&\gvec{x}_t \in\mathbb{R}^m,\label{eqn:SSModelState}\\
\gvec{y}_{t}&= \gvec{h}\brackets{\gvec{x}_t}+\gvec{v}_t,
&\gvec{y}_t \in\mathbb{R}^n.\label{eqn:SSModelObs}
\end{align}
\end{subequations}

We consider a case where at least some of the state variables can be estimated from $\gvec{y}_t$, which is related to the notion of observablity, typically used in the context of deterministic systems \cite[Ch. 3]{bar2004estimation}. {In particular, we use the term {\em fully observable} to denote measurement models where $\gvec{y}_t$ is affected by all variables in $\gvec{x}_t$, and all variables in state $\gvec{x}_t$ can be recovered from $\gvec{y}_t$ (i.e., the mapping  $\gvec{h}(\cdot)$ is injective). {\em partially observable} for models in which some of the entries of $\gvec{x}_t$ cannot be recovered from $\gvec{y}_t$ (though they may be dependent on $\{\gvec{y}_{\tau}\}_{\tau \leq t}$).} 
That is, we examine both the fully observable case  and the partially observable setting, where in the latter a single $\gvec{y}_t$, can be used to recover only a subset of $p \leq m$ variables in $\gvec{x}_t$, denoted as the $p \times 1$ vector $ \gvec{P}\gvec{x}_t$, with $\gvec{P}$ being a $p \times m$ selection matrix. We henceforth focus our on the partially observable setting as it also includes the fully observable setting by writing  $\gvec{P}=\gvec{I}$ and $p=m$.

Our goal is to develop a filtering algorithm for real-time state estimation, i.e., for the recovery of $\gvec{x}_t$ 
from $\{\gvec{y}_\tau\}_{\tau\leq t}$ for each time instance $t$ \cite{durbin2002simple}.
This algorithm should work effectively
in both fully or partially observable \ac{ss} models, while we assume that one has knowledge on which state variables are observable, i.e., $\gvec{P}$ is known. The performance of a given estimator (obtained by the filtering approach) $\hat{\gvec{x}}_t$ is measured using the \ac{mse},  which is defined as  $\mathbb{E}\{\|\hat{\gvec{x}}_t - {\gvec{x}}_t\|^2\}$.

While various methods have been proposed for tracking in  \ac{ss} models, our setting is associated with several  challenges: 
\begin{enumerate}[label={
C.\arabic*}]
\item \label{itm:Dist} The distribution of the noises, $\gvec{e}_t$ and $\gvec{v}_t$ in \eqref{eqn:SSModel}, is unknown and may be non-Gaussian as, e.g., stochasticity in visual data is often non-Gaussian.
\item \label{itm:Approx} The available state-evolution function, $\gvec{f}(\cdot)$, may be mismatched, e.g., obtained via a first-order linear approximation of complex physical dynamics, as is often the case in navigation and localization tasks~\cite[Ch. 6]{bar2004estimation}.
\item \label{itm:high-dimensional} The observations are high-dimensional ($n\gg m$), leading to high complexity and affecting real-time applicability.
\item \label{itm:unknown_h}  The sensing function $\gvec{h}(\cdot)$  is unknown and possibly analytically intractable.
\end{enumerate}

To cope with the various unknown characteristics of \eqref{eqn:SSModel}, we are given access to
a labeled data set comprised of $D$ trajectories of length $T$ of paired observations and states, 
\begin{equation}
\label{eqn:DataSet}
    \mySet{D} \triangleq \left\{\left\{\gvec{x}_t^{(d)}, \gvec{y}_t^{(d)}\right\}_{t=1}^T \right\}_{d=1}^{D}.
\end{equation} 
Our proposed algorithm for tackling \ref{itm:Dist}-\ref{itm:unknown_h} is detailed in Section~\ref{sec:Learned_Kalman_Filtering_in_the_Latent_Space}. Our design follows  model-based deep-learning methodology \cite{shlezinger2022model,shlezinger2020model}, where deep-learning tools are used  to augment and empower model-based algorithms rather than replace them. 
Our method builds upon the \acl{kn} architecture of \cite{revach2022kalmannet}, which augments the classic \ac{ekf}, as briefly recalled in the next subsection. 

 
\subsection{\ac{ekf} and \acl{kn}
}\label{ssec:Data-Driven_Filtering_with_KalmanNet}
 
Various model-based  filters have been developed tracking in \ac{ss} models  (see, e.g. \cite[Ch. 10]{durbin2012time}). One of the most common algorithms, which is suitable when the noises are Gaussian and the \ac{ss} model is fully known, i.e., in the absence of Challenges \ref{itm:Dist}-\ref{itm:unknown_h}, is the \ac{ekf} \cite{gruber1967approach}. The \ac{ekf} follows the operation of the \ac{kf}, combining  prediction based on the previous estimate with an update based on the current observation, while extending it to nonlinear \ac{ss} models. 

In particular, the \ac{ekf} first predicts the next state and observation based on  $\hat{\gvec{x}}_{t-1}$ via
\begin{equation}
\label{eqn:Pred}
    \hat{\gvec{x}}_{t|t-1} = \gvec{f}(\hat{\gvec{x}}_{t-1}); \quad \hat{\gvec{y}}_{t|t-1} = \gvec{h}(\hat{\gvec{x}}_{t|t-1}).
\end{equation}
Then, the initial prediction is updated with a  matrix $\Kgain_{t}$, known  as the Kalman gain, which dictates the balance between relying on the state evolution function $f(\cdot)$ through \eqref{eqn:Pred} and the current observation $\gvec{y}_t$. The estimate is computed as
\begin{equation}\label{eqn:EKFUpdate}
 \hat{\gvec{x}}_{t} = \Kgain_{t}\cdot \Delta \gvec{y}_t +  \hat{\gvec{x}}_{t|t-1}; \qquad \Delta \gvec{y}_t \triangleq \gvec{y}_t -  \hat{\gvec{y}}_{t|t-1}. 
\end{equation}
The Kalman gain $\Kgain_{t}$  is calculated via 
\begin{equation}
\label{eqn:kalman_gain_computaion}
    \Kgain_{t} = \gvec{\hat{\gvec{\Sigma}}}_{t|t-1}\cdot\gvec{\hat{H}}_t^\top\cdot \gvec{\hat{S}}_{t|t-1}^{-1},
\end{equation}
where $\gvec{\hat{\Sigma}}_{t|t-1}$ and $\gvec{\hat{S}}_{t|t-1}$ are the {covariance matrices of the state prediction $\hat{\gvec{x}}_{t|t-1}$ and observation prediction $\hat{\gvec{y}}_{t|t-1}$, respectively. These matrices are calculated via }
\begin{equation}
\label{eqn:state_covariance_computaion}
\gvec{\hat{\gvec{\Sigma}}}_{t|t-1}=\hat{\gvec{F}}_{t}\cdot\hat{\gvec{\Sigma}}_{t-1}\cdot\hat{\gvec{F}}_t^\top+\gvec{Q},
\end{equation}
\begin{equation}
\label{eqn:obs_covariance_computaion}
\gvec{\hat{S}}_{t|t-1}=\hat{\gvec{H}}_t\cdot\gvec{\tilde{\gvec{\Sigma}}}_{t|t-1}\cdot\hat{\gvec{H}}_t^\top+\gvec{R},
\end{equation}
where $\gvec{Q}$ and $\gvec{R}$ are the known covariance matrices of $\gvec{e}_t$ and $\gvec{v}_t$, respectively. The matrices $\hat{\gvec{F}}_{t}$ and $\hat{\gvec{H}}_{t}$ are instantaneous linearizations of $\gvec{f}(\cdot)$ and $\gvec{h}(\cdot)$, respectively, obtained using their Jacobian matrices  evaluated at $\hat{\gvec{x}}_{t-1}$ and $\hat{\gvec{x}}_{t|t-1}$ (see \cite[Ch. 10]{durbin2012time}), i.e., 
\begin{align}
\label{eqn:Jacobians}
\hat{\gvec{F}}_{t}=\nabla_{\gvec{x}}\gvec{f}(\hat{\gvec{x}}_{t-1}); \quad 
\hat{\gvec{H}}_{t}=\nabla_{\gvec{x}}\gvec{h}(\hat{\gvec{x}}_{t|t-1}).
\end{align} 

Challenges \ref{itm:Dist}-\ref{itm:unknown_h} notably limit the applicability of the \ac{ekf} for the setup detailed in the previous subsection. 
\acl{kn} proposed at \cite{revach2022kalmannet} is designed to leverage data as in \eqref{eqn:DataSet} to tackle Challenges \ref{itm:Dist} and \ref{itm:Approx} (but not \ref{itm:high-dimensional}-\ref{itm:unknown_h}). In particular, \acl{kn} builds on the insight that the missing and mismatched domain knowledge of the noise statistics and the linear approximations are encapsulated in the computation of the Kalman gain $\Kgain_{t}$ \eqref{eqn:kalman_gain_computaion}. Consequently, it augments the \ac{ekf} with a deep-learning component by replacing the computation of the Kalman gain with an \ac{rnn}, while preserving the filtering operation via \eqref{eqn:Pred}-\eqref{eqn:EKFUpdate}. 
By doing so, \acl{kn} converts the \ac{ekf} into a trainable discriminative model~\cite{shlezinger2022discriminative}, where the data $\mySet{D}$ is used to directly learn the Kalman gain, bypassing the need to enforce any model over the noise statistics and able to handle domain knowledge mismatches as in challenges \ref{itm:Dist}-\ref{itm:Approx}. Moreover, \acl{kn} preserves the interoperability of the \ac{kf}, while being operable in partially known \ac{ss} models; therefore it allows to deduce uncertainty as shown in \cite{klein2022uncertainty}, and is amenable to training in an unsupervised manner~\cite{revach2022unsupervised}. 

Despite the ability of the \acl{kn} architecture of \cite{revach2022kalmannet} to learn from data to cope with Challenges~\ref{itm:Dist} and \ref{itm:Approx}, it is not suitable to be applied in our setting under \ref{itm:high-dimensional}-\ref{itm:unknown_h}. In particular, the high dimension of the observations notably increases the complexity of its Kalman gain \ac{rnn} and the resulting filter. Moreover, \acl{kn} requires knowledge of $\gvec{h}(\cdot)$, which is not analytically available in the current setting. This motivates the derivation of the proposed \name~in the sequel. 


\section{\name}
\label{sec:Learned_Kalman_Filtering_in_the_Latent_Space}

In this section, we present the proposed \name~algorithm, which tackles Challenges \ref{itm:Dist}-\ref{itm:unknown_h}. 
Our derivation of \name~is presented in a step-by-step manner, where each step tackles an additional challenging aspect, while builds upon its preceding stages.

As noted in Subsection~\ref{ssec:Data-Driven_Filtering_with_KalmanNet}, the main added challenges considered here as compared to the setting for which \acl{kn} is formulated in \cite{revach2022kalmannet} are associated with the observation model in \eqref{eqn:SSModelObs}, i.e., Challenges \ref{itm:high-dimensional} and \ref{itm:unknown_h}. Therefore, our first step, detailed in Subsection~\ref{ssec:Step_1-Instantaneous_Estimate}, considers an instantaneous estimation setting based solely on the observations model.
The second step, described in Subsection~\ref{ssec:Step_2-Kalman_in_Latent}, incorporates the state evolution function \eqref{eqn:SSModelState} by adding a prior as an additional input, which accounts for temporal correlation and partial observability. Then, we unite the instantaneous estimate with the model-based \ac{ekf} for tracking in Step 3 (Subsection~\ref{ssec:Step_3-Joint_Tracking_and_Encoding}) to face Challenges \ref{itm:high-dimensional}-\ref{itm:unknown_h}.
Our fourth step, detailed in Subsection~\ref{ssec:Step_4-Jointly_Learned_Tracking_and_Encoding}, incorporates Challenges \ref{itm:Dist} and \ref{itm:Approx} by converting the joint instantaneous estimate and tracking algorithm into \name, by replacing EKF with KalmanNet. \name \ is learned as a trainable discriminative model,  where \acl{dd} tracking is done based on jointly learned latent features. We conclude our derivation with a discussion in Subsection~\ref{ssec:Discussion}.

\subsection{Step 1 - Instantaneous Estimate}
\label{ssec:Step_1-Instantaneous_Estimate}
We begin by considering the observations model \eqref{eqn:SSModelObs} solely. The resulting task boils down to the instantaneous estimation of (the observable entries of) $\gvec{x}_t$ from the observed $\gvec{y}_t$. The fact that the observation model is unknown (\ref{itm:unknown_h}) and high-dimensional (\ref{itm:high-dimensional}), combined with the availability of labeled data \eqref{eqn:DataSet}, motivates the usage of \acp{dnn}. 
A natural approach here is to use a regression \ac{dnn} with parameters $\Encparams$, denoted   $\gvec{g}^{\rm e}_{\Encparams}:\mathbb{R}^n\mapsto \mathbb{R}^p$, 
and train it to map $\gvec{y}_t$ into an estimate of the observable state variables $\gvec{P}\gvec{x}_t$. 

When properly trained with sufficient data,  regression \acp{dnn} are often capable of learning to provide 
reliable estimates from high-dimensional data with complex statistical models \cite[Ch. 13]{goodfellow2016deep}. 
In particular, the \ac{dnn} parameters $\Encparams$ can be learned in a supervised manner via gradient-based optimization, e.g., \ac{sgd} and its variants. We adopt the regularized $\ell_2$ norm \ac{mse} loss, which, for a given data set $\mySet{D}$, is computed as: 
\begin{equation}
\label{eqn:loss_function_encoder}
\mySet{L}_{\mySet{D}}^{\rm e}(\Encparams)\!=\!\frac{1}{|\mySet{D}|  T}\sum_{d=1}^{|\mySet{D}|}  \sum_{t=1}^T \norm{  \gvec{g}_{\Encparams}^{\rm e}\big(\gvec{y}_t^{(d)}\big)\!-\!\gvec{P}\gvec{x}^{(d)}_t}^2 \!+\! \lambda \norm{\Encparams}^2,
\end{equation}
where $\lambda > 0$ is a regularization coefficient. 

The resulting instantaneous estimation system represents a straightforward \acl{dd} approach to tackle the challenges associated with the observation model \eqref{eqn:SSModelObs}. However, it does not account for the temporal correlation induced by the state evolution model \eqref{eqn:SSModelState}.  Moreover, the full reliance on black box deep-learning architectures implies that the resulting system is sensitive to generalization problems and the model can easily over-fit to the training set.

\subsection{Step 2 - Incorporating the Evolution Model}
\label{ssec:Step_2-Kalman_in_Latent}
 The instantaneous estimator is oblivious to the partially known state evolution model in \eqref{eqn:SSModelState}. Therefore, by 
integrating  the model in \eqref{eqn:SSModelState}, i.e., the (possibly approximated) state evolution function $\gvec{f}(\cdot)$, 
one can potentially improve the estimation of the observable state variables provided by $\gvec{g}^{\rm e}_{\Encparams}(\cdot)$. Our rationale stems from viewing the inference task carried out by the \ac{dnn}, i.e., recovering $\gvec{P}\gvec{x}_t$ from $\gvec{y}_t$, as solving a non-convex optimization problem (see \cite{shlezinger2022model}). While tackling non-convex optimization is in general highly challenging, it can be greatly facilitated by providing a good initial guess, which hopefully lies in the proximity of the global optimum \cite{park2017general}. 

\begin{figure}
\centering
\includegraphics[width=\columnwidth]{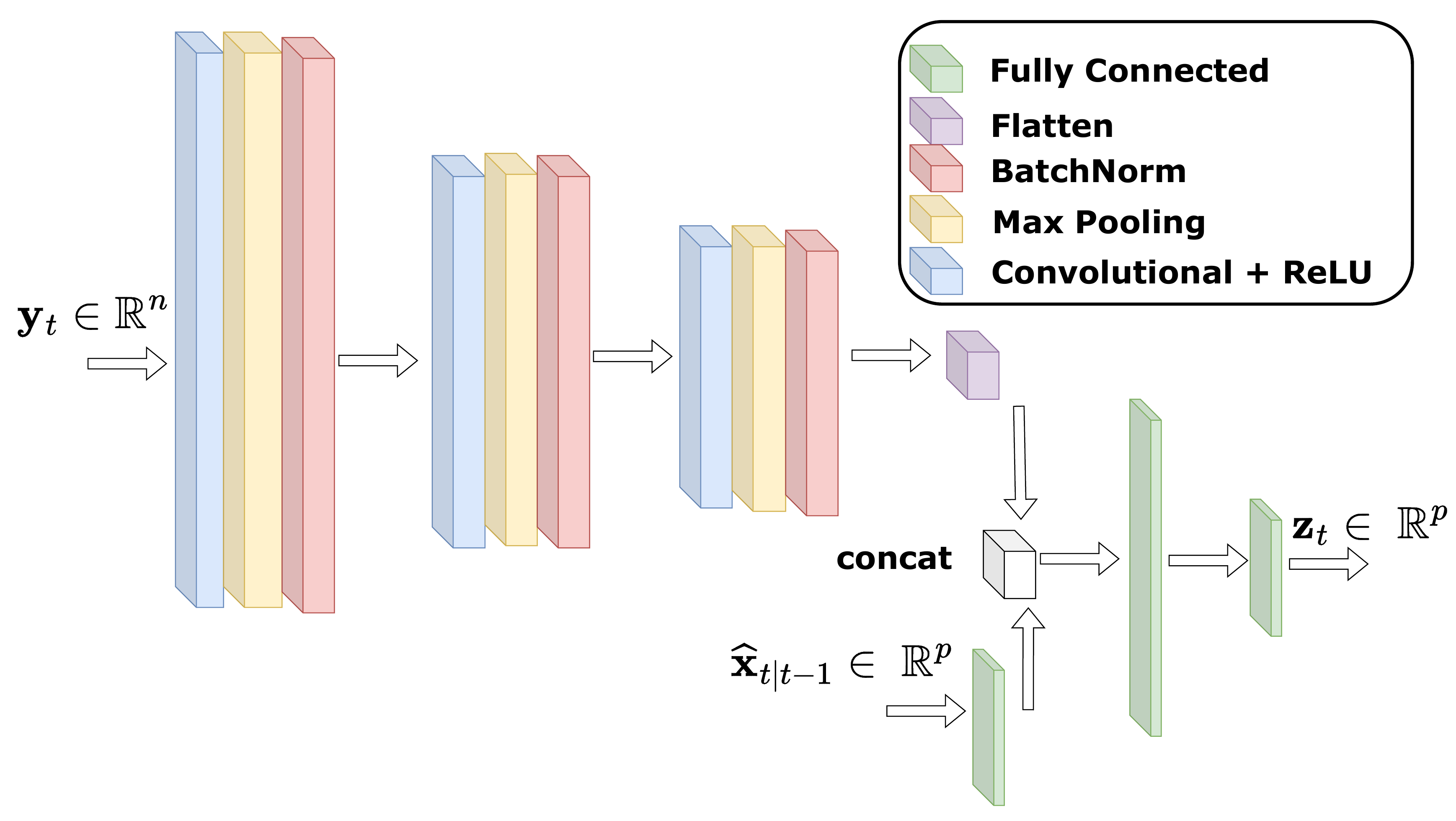}
\caption{Illustration of an encoder with prior implemented using a \ac{cnn}, following the implementation utilized in the numerical study in Section~\ref{sec:Empirical_Study}.} 
\vspace{0.2cm}
\label{fig:Encoder with prior block diagram}
\end{figure}

To incorporate the state evolution as a form of a prior, we apply $\gvec{f}(\cdot)$ to the previous estimate, $\hat{\gvec{x}}_{t-1}$. The previous estimate $\hat{\gvec{x}}_{t-1}$ can be obtained from the previous encoder output, i.e., $\gvec{g}_{\Encparams}^{\rm e}\big(\gvec{y}_{t-1}\big)$, while fused with an estimate of the unobservable variables, that can be provided as an initial guess and improve the instantaneous estimate by using the temporal correlation.
 An example of the resulting high-level architecture, where the prior prediction $\hat{\gvec{x}}_{t|t-1}$ is provided to the \ac{dnn}-based instantaneous estimate (implemented as a \ac{cnn}) as additional features, is illustrated in Fig.~\ref{fig:Encoder with prior block diagram}. The \ac{dnn}-based estimator now takes two multivariate inputs, $\hat{\gvec{x}}_{t|t-1}$ and $\gvec{y}_t$, and is thus trained for mapping $\mathbb{R}^n\times \mathbb{R}^m$ into $\mathbb{R}^p$. Namely, here 
\begin{equation}
\label{eqn:NewLatent}
    \gvec{z}_t = \gvec{g}^{\rm e}_{\Encparams}(\gvec{y}_t, \hat{\gvec{x}}_{t|t-1}).
\end{equation}

The encoder is trained using the regularized \ac{mse} loss as in \eqref{eqn:loss_function_encoder}. The incorporation of past outputs as a prior enables leveraging domain knowledge in the sense of the state evolution model, thus guiding the learning procedure towards a more desirable solution. 
Yet, it also impacts the stability of the training procedure, as we have empirically observed. Nonetheless, the learning procedure can be facilitated by exploiting the interpretability of the architecture, building upon the ability to view the prior   $\hat{\gvec{x}}_{t|t-1}$ as a noisy version of the desired state. 
One can thus set the prior during training to be the ground truth state with added noise, while choosing the noise magnitude based on an ablation study, as done in our numerical study reported in Section~\ref{sec:Empirical_Study}. 
Here, care should be taken not to push the encoder to learn solely from the observation when the noise is too large, while not relying only upon the prior where the noise is too small. 

\subsection{Step 3 - Joint Instantaneous Estimation and Tracking}
\label{ssec:Step_3-Joint_Tracking_and_Encoding}
\begin{figure}
\centering
\includegraphics[width=\columnwidth]{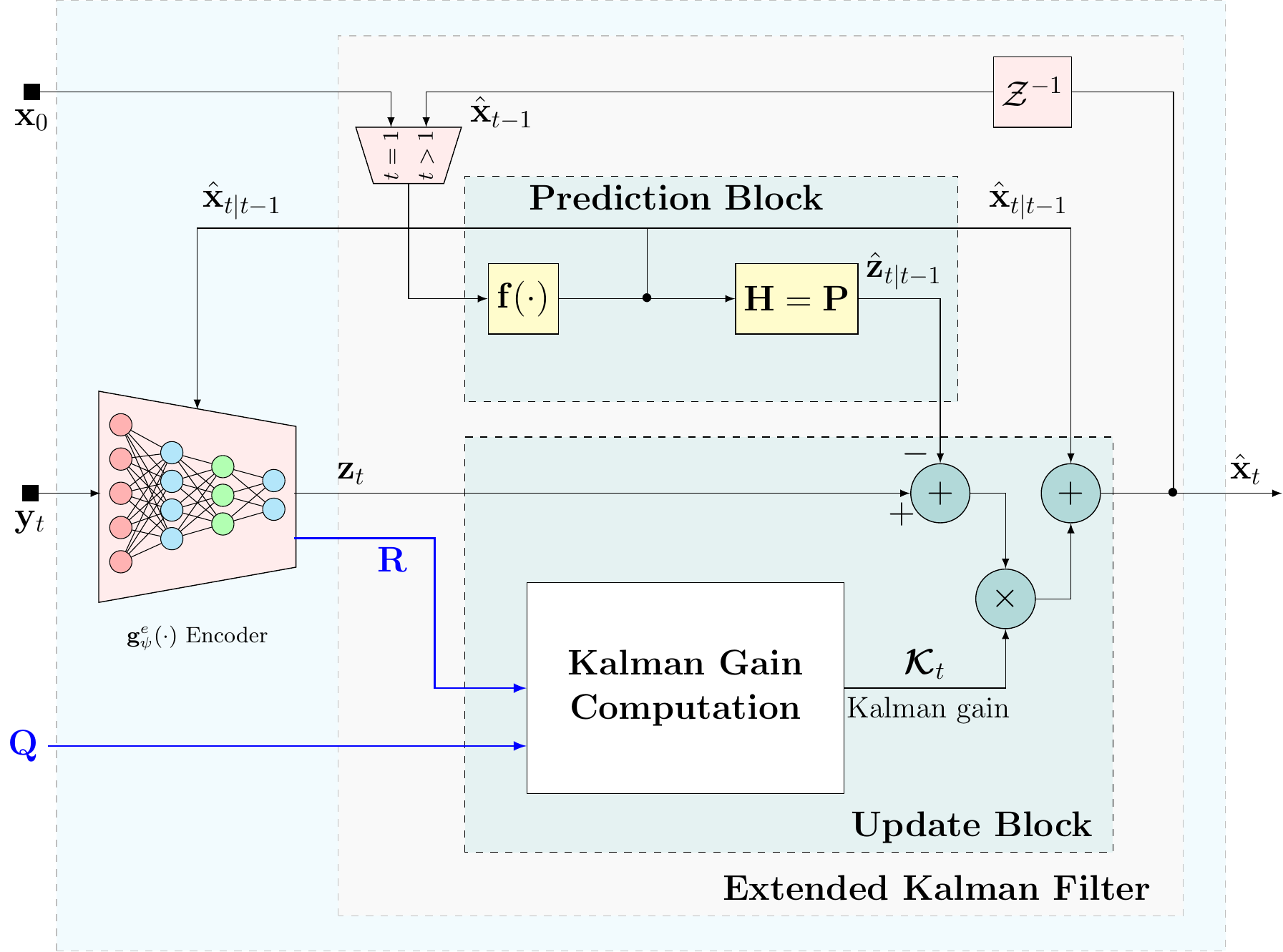}
\caption{Encoder with prior and EKF in cascade block diagram. 
}
\label{fig:Encoder with prior knowledge and EKF in cascade block diagram}
\end{figure}

Next, we assume that the relationship between the estimator output and the observable state variables can be represented as obeying a  Gaussian distribution. This approach allows tracking in the latent space, without accounting Challenges \ref{itm:Dist} and \ref{itm:Approx}. In such cases, one can account for the temporal correlation by applying an \ac{ekf} 
in cascade with the  pretrained \ac{dnn} encoder of Step 2. The rationale here is to assume that the \ac{dnn} is properly trained such that its estimate approaches the minimal \ac{mse} estimator of $\gvec{P}\gvec{x}_t$ from $\gvec{y}_t$. In such cases, the \ac{dnn} output can be approximated as obeying
\begin{equation}
    \label{eqn:SSModelEst}
        \gvec{z}_t = \gvec{g}^{\rm e}_{\Encparams}(\gvec{y}_t) \approx  \gvec{P}\gvec{x}_t + \tilde{\gvec{v}}_t,
\end{equation}
where $\tilde{\gvec{v}}_t$ is zero-mean and mutually independent of $\gvec{x}_t$. If $\tilde{\gvec{v}}_t$ is also Gaussian and temporally independent, then \eqref{eqn:SSModelState} along with \eqref{eqn:SSModelEst} represent a (possibly nonlinear) Gaussian \ac{ss} model, from which $\gvec{x}_t$ can be tracked using the \ac{ekf}. The second-order moment of $\tilde{\gvec{v}}_t$, which is necessary for the Kalman gain computation \eqref{eqn:kalman_gain_computaion}, can be estimated from the validation error $\gvec{R}$ of the \ac{dnn} encoder. {The measurement matrix $\hat{\gvec{H}}_{t}$ in this setting is set to $\hat{\gvec{H}}_{t} = \gvec{P}$.} The  system is illustrated in Fig.~\ref{fig:Encoder with prior knowledge and EKF in cascade block diagram}. 

To apply the \ac{ekf} in latent space while treating \eqref{eqn:SSModelEst} as the observation model, one should have knowledge of the distribution of the state noise  $\gvec{e}_t$, i.e., the matrix  $\gvec{Q}$. This can be  estimated from the data $\mySet{D}$. For instance, one can tune the dynamic noise variance to optimize performance by, e.g., assuming that $\gvec{Q}$ is a scaled identity matrix and employing grid search to identify the variance parameter which yields the best performance on the available data. Alternatively, one can incorporate parametric estimation mechanisms, e.g., expectation maximization iterations, into  the \ac{ekf}~\cite{gannot2008kalman}.  

%
%

The proposed cascaded operation allows utilizing a \ac{dnn} to cope with the challenging observations model while  systematically incorporating the state evolution model. This is achieved by separating instantaneous estimation from the tracking task, where the temporal correlation is exploited.  Jointly treating the instantaneous estimate task along with its subsequent tracking allows for improving the overall performance, as shown in Section~\ref{sec:Empirical_Study}. Nonetheless, the fact that an \ac{ekf} is utilized implies that the \ac{ss} described via \eqref{eqn:SSModelState} and \eqref{eqn:SSModelEst} is inherently assumed to be fully known and Gaussian. This is not necessarily the case here,  not only due to Challenges \ref{itm:Dist} and \ref{itm:Approx}, but also since there is no guarantee that the \ac{dnn} estimation error $\tilde{\gvec{v}}_t$  is indeed Gaussian. This motivates our final step, which formulates \name.

\subsection{Step 4 - \name}
\label{ssec:Step_4-Jointly_Learned_Tracking_and_Encoding}
The system detailed in Step 3 builds upon the insight that the relationship between the latent $\gvec{z}_t$ and the state $\gvec{x}_t$ obeys an (approximated) \ac{ss} model given by \eqref{eqn:SSModelState} and \eqref{eqn:SSModelEst}. We conclude our design by accounting for Challenges \ref{itm:Dist} and \ref{itm:Approx}, and the fact that the error term in \eqref{eqn:SSModelEst} is likely to obey an  unknown distribution. This motivates using \acl{kn} instead of the \ac{ekf}, which is particularly  suitable for filtering in such settings, and bypasses the need to impose a specific distribution on the noise terms in the \ac{ss} model. The resulting algorithm, encompassing the architecture and its training procedure detailed next, is coined {\em \name}.

\vspace{0.1cm}

{\bf Architecture:} 
To formulate the system operation, we let $\KNparams$ be the internal \ac{rnn} parameters of \acl{kn}, which implements a mapping $\gvec{g}^{\rm f}_{\KNparams}:\mathbb{R}^p\mapsto \mathbb{R}^m$ with state-evolution function $\gvec{f}(\cdot)$ and  observation function given by $\gvec{h}(\gvec{x})=\gvec{P}\gvec{x}$. 
As detailed in Subsection~\ref{ssec:Data-Driven_Filtering_with_KalmanNet}, \acl{kn} uses the previous estimate $\hat{\gvec{x}}_{t-1}$ to predict the next state as $\hat{\gvec{x}}_{t|t-1} = \gvec{f}\big(\hat{\gvec{x}}_{t-1}\big)$. This prediction is then used as the prior provided encoder of Step 2, producing the latent $\gvec{z}_t$ via~\eqref{eqn:NewLatent}. 
The estimate of $\gvec{x}_t$ is written as
\begin{equation}
    \hat{\gvec{x}}_t = \gvec{g}^{\rm f}_{\KNparams}(\gvec{z}_t).
    \label{eqn:LatentFilter}
\end{equation}

%
%
\begin{figure}
\includegraphics[width=1\columnwidth]{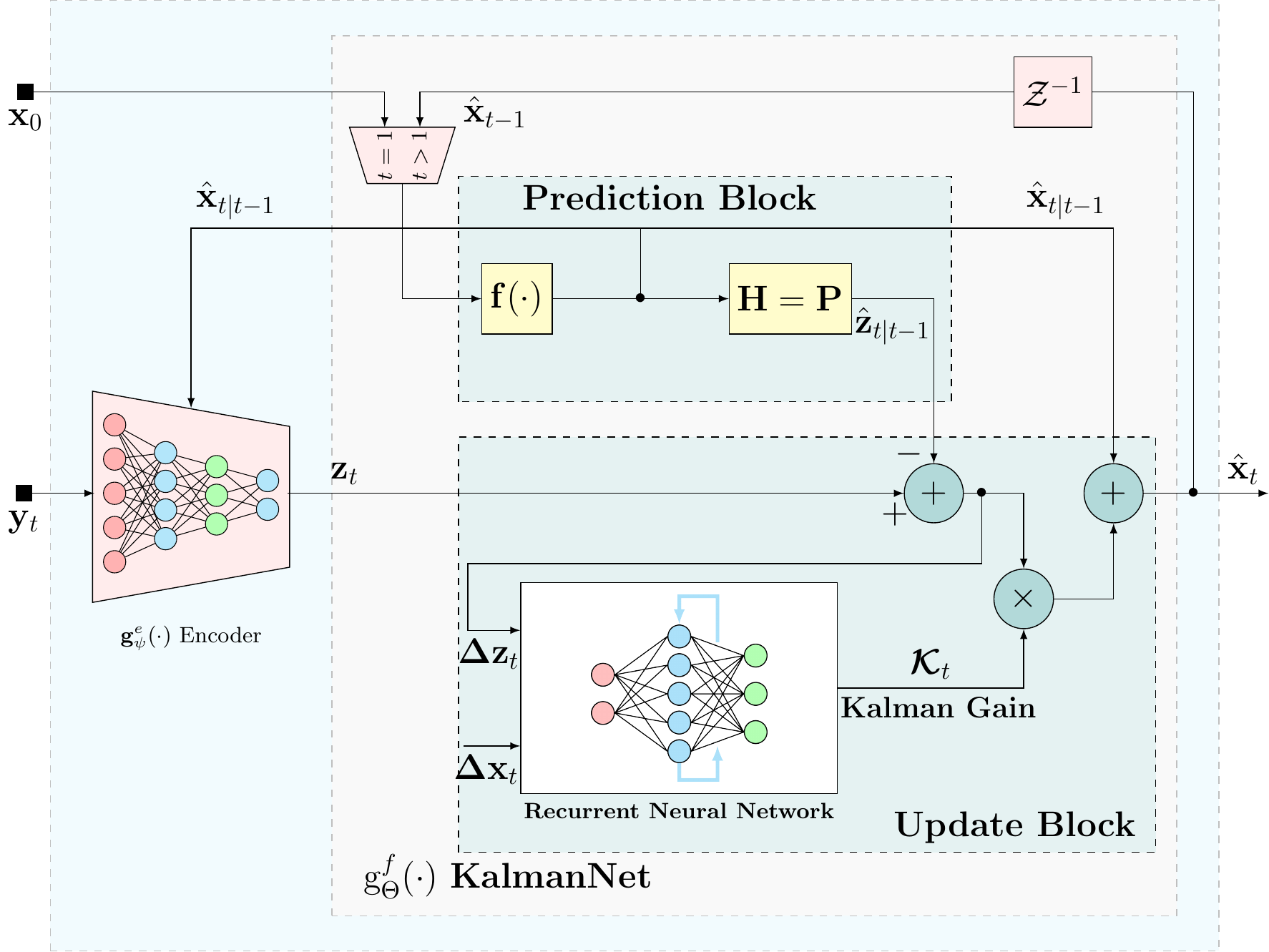}
\caption{\name~block diagram.}
\label{fig:Learned Kalman filtering in latent space block diagram}
\end{figure}

The resulting architecture is illustrated in Fig.~\ref{fig:Learned Kalman filtering in latent space block diagram}, where the two modules, the encoder and \acl{kn}, aid one another by providing a low-dimensional latent representation (by the encoder), and a prior for obtaining the latent (by \acl{kn}). {Once trained, the estimation procedure on each time step, during inference, is summarized as Algorithm~\ref{alg:LatKN}.}

  \begin{algorithm}
    \caption{\name~Inference}
    \label{alg:LatKN} 
    \SetKwInOut{Initialization}{Init}
    \Initialization{Trained encoder $\Encparams$; Trained \acl{kn} $\KNparams$}
    \SetKwInOut{Input}{Input} 
    \Input{Observations $\gvec{y}_t$; previous estimate $\hat{\gvec{x}}_{t-1}$}
    {
      Predict $ \hat{\gvec{x}}_{t|t-1} = \gvec{f}(\hat{\gvec{x}}_{t-1})$ \;
      Predict $\hat{\gvec{z}}_{t|t-1} = \gvec{P}\hat{\gvec{x}}_{t|t-1}$\;
      Encode observations via 
      $\gvec{z}_t = \gvec{g}^{\rm e}_{\Encparams}(\gvec{y}_t, \hat{\gvec{x}}_{t|t-1})$\;
      Apply the \ac{rnn}  $\KNparams$ to compute Kalman gain $\Kgain_t$\;
      Estimate via $\hat{\gvec{x}}_{t} = \gvec{g}^{\rm f}_{\KNparams}(\gvec{z}_t) = \Kgain_{t}\cdot (\gvec{z}_t -\hat{\gvec{z}}_{t|t-1} ) +  \hat{\gvec{x}}_{t|t-1}$\;
  }\KwRet{$\hat{\gvec{x}}_{t}$}
\end{algorithm}

{\bf Training} 
The proposed architecture is a concatenation of two modules: A \ac{dnn} estimator $\gvec{g}^{\rm e}_{\Encparams}(\cdot)$ and \acl{kn} $\gvec{g}^{\rm f}_{\KNparams}(\cdot)$. Both are differentiable \cite{revach2022kalmannet}, allowing the overall architecture, parameterized by $(\KNparams,\Encparams)$, to be trained end-to-end as a discriminative model \cite{shlezinger2022discriminative}. We  use the $\ell_2$ regularized \ac{mse} loss, which for a given data set $\mySet{D}$ is evaluated as
\begin{align}
\mySet{L}_{\mySet{D}}(\KNparams,\Encparams)=&\frac{1}{|\mySet{D}|T}\sum_{d=1}^{|\mySet{D}|}  \sum_{t=1}^\top \mySet{L}^{(d)}_t(\KNparams,\Encparams) \notag \\ & +\lambda_1 \norm{\KNparams}^2+\lambda_2 \norm{\Encparams}^2,
\label{eqn:loss_function}
\end{align}
where $\lambda_1,\lambda_2>$ 0 are  regularization coefficients. The loss term for each time step in a given trajectory of \eqref{eqn:loss_function} is computed as
\begin{equation}
\label{eqn:loss_function_time_step}
\mySet{L}^{(d)}_t(\KNparams,\Encparams) = \norm {\hat{\gvec{x}}^{(d)}_t -\gvec{x}^{(d)}_t}^2,   
\end{equation}
where
\begin{align} 
\hat{\gvec{x}}^{(d)}_t &= \gvec{g}_{\KNparams}^{\rm f}\big( \gvec{g}_{\Encparams}^{\rm e}\big(\gvec{y}_t^{(d)}, \gvec{f}\big(\hat{\gvec{x}}^{(d)}_{t-1}\big)\big)\big) \notag\\
&= \hat{\gvec{x}}^{(d)}_{t|t-1} + \Kgain_{t}\cdot (\gvec{z}_t -\hat{\gvec{z}}_{t|t-1} ). 
\label{eqn:state estimation equation}
\end{align}

The loss measure in \eqref{eqn:loss_function} builds upon the ability to backpropagate the loss to the computation of the Kalman gain $\Kgain_{t}$ \cite{xu2021ekfnet}. In particular, One can obtain the loss gradient of a given trajectory $d$ in given time step $t$ with respect to the Kalman gain from the output $\hat{\gvec{x}}^{(d)}_t$ of \name~since
\begin{align}
\frac{\partial \mySet{L}^{(d)}_t(\KNparams,\Encparams)}{\partial \Kgain_{t}} &= \frac{\partial \norm {\Kgain_{t} \Delta \gvec{z}_t - \Delta \gvec{x}_t}^2}{\partial \Kgain_{t}} \notag \\
&=2(\Kgain_{t} \Delta \gvec{z}_t - \Delta \gvec{x}_t)\cdot \Delta \gvec{z}_t ^\top,
\label{eqn:loss_gradient}
\end{align}
where $\Delta \gvec{x}_t=\gvec{x}^{(d)}_t - \hat{\gvec{x}}^{(d)}_{t|t-1}$. The gradient computation in \eqref{eqn:loss_gradient} indicates that one can learn the computation of the Kalman gain by training \name~end-to-end. This allows training the overall filtering system, including both the latent encoding and its tracking into the state without having to externally provide ground truth values of the Kalman gain or of the latent features for training purposes.
The fact that the  \ac{mse} loss in \eqref{eqn:loss_function_time_step} is computed based on the output of \acl{kn} rather than that of $\gvec{g}^{\rm e}_{\Encparams}(\cdot)$, implies that the latter will not necessarily learn to estimate the observable state variables, as in when training via \eqref{eqn:loss_function_encoder}. Instead, it is trained to encode the high-dimensional observations $\gvec{y}_t$ (along with the prior $\hat{\gvec{x}}_{t|t-1}$) into {\em latent features}, from which \acl{kn} can most reliably recover the state. For this reason, we coin the algorithm {\em \name}. 

\name~enables joint learning of $(\KNparams,\Encparams)$ via gradient-based optimization, e.g., \ac{sgd} and its variants. However, carrying this out in practice can be challenging and often unstable, as the learning procedure needs to simultaneously tune the latent representation and the corresponding Kalman gain computation. Nonetheless, the fact that the architecture is decomposable into distinct trainable building blocks with concrete tasks facilitates training via alternating optimization. This is achieved by iteratively optimizing the filter $\KNparams$ while freezing $\Encparams$, followed by training of the latent representation $\Encparams$ which best fits the filter with 
fixed
weights $\KNparams$ based on \eqref{eqn:loss_function}. Additionally, one can initially train the observable variables of the encoder module separately, via the regularized $\ell_2$ norm loss \eqref{eqn:loss_function_encoder}. This form of modular training~\cite{raviv2022online} constitutes a warm start which is empirically shown to facilitate learning. The resulting procedure is summarized as Algorithm~\ref{alg:Alternating}.

  \begin{algorithm}
    \caption{\name~Alternating Training}
    \label{alg:Alternating} 
    \SetKwInOut{Initialization}{Init}
    \Initialization{Fix learning rates $\mu_1,\mu_2>0$ and  epochs $i_{\max}$}
    \SetKwInOut{Input}{Input} 
    \Input{Training set  $\mySet{D}$}  
    {
        \nonl\texttt{Warm start:}\\
        \For{$i = 0, 1, \ldots, i_{\max}-1$}{%
                    Randomly divide  $\mySet{D}$ into $Q$ batches $\{\mySet{D}_q\}_{q=1}^Q$\;
                    \For{$q = 1, \ldots, Q$}{
                    Compute batch loss $\mathcal{L}_{\mySet{D}_q}^{\rm e}(\Encparams)$  by \eqref{eqn:loss_function_encoder}\;
                    Update  $\Encparams\leftarrow \Encparams - \mu_1\nabla_{\Encparams}\mathcal{L}^{\rm e}_{\mathcal{D}_q}(\Encparams)$\;
                    }
        }
        \nonl\texttt{Alternating minimization:}\\
        \For{$i = 0, 1, \ldots, i_{\max}-1$}{%
                    Randomly divide  $\mySet{D}$ into $Q$ batches $\{\mySet{D}_q\}_{q=1}^Q$\;
                    \For{$q = 1, \ldots, Q$}{
                    Compute batch loss $\mathcal{L}_{\mySet{D}_q}(\KNparams,\Encparams)$  by \eqref{eqn:loss_function}\;
                    Update  $\KNparams\leftarrow \KNparams - \mu_2\nabla_{\KNparams}\mathcal{L}_{\mathcal{D}_q}(\KNparams,\Encparams)$\;
                    }
                    \For{$q = 1, \ldots, Q$}{
                    Compute batch loss $\mathcal{L}_{\mySet{D}_q}(\KNparams,\Encparams)$  by \eqref{eqn:loss_function}\;
                      Update  $\Encparams\leftarrow \Encparams - \mu_1\nabla_{\Encparams}\mathcal{L}_{\mathcal{D}_q}(\KNparams,\Encparams)$\;
                 }                    
        }
        \KwRet{($\KNparams,\Encparams)$}
  }
  
\end{algorithm}

\subsection{Discussion}
\label{ssec:Discussion}
The proposed \name~is designed to tackle the challenges of tracking from complex high-dimensional observations. It leverages data to enable reliable tracking, overcoming the missing knowledge of the sensing function and the noise statistics. \name~is derived in gradual steps obtained from pinpointing the specific challenges associated with the filtering problem detailed in Subsection~\ref{ssec:Problem_Formulation}. In particular, the usage of a \ac{dnn} trained in a supervised manner for coping with the complex observations model in Step 1 is a straightforward approach for instantaneous estimations. Its cascading with an \ac{ekf} is a natural extension for incorporating temporal correlation, and a similar approach of applying an \ac{ekf} to \acl{dd} extracted features was also proposed in previous works, e.g., \cite{coskun2017long, zhou2020kfnet}. 
However, the usage of the evolution model $\gvec{f}(\cdot)$  in Step 2 for improving instantaneous estimate due to temporal correlation; replacing the \ac{ekf} with the trainable \acl{kn} in Step 4; and the formulation of a suitable training procedure which encourages both modules to facilitate their counterpart's operation in tracking, are novel aspects of our design. These components are particularly tailored to cope with the challenging partially known \ac{ss} model in \eqref{eqn:SSModel}, without enforcing a model on the noise statistics and the observations function, and while being geared towards real-time applications with low-latency inference demands. 

Compared with the preliminary findings of this research reported in \cite{LatentICASSP}, the \name~algorithm presented here is not restricted to using instantaneous estimators for latent feature extraction, and can in fact learn to contribute to the latent state encoding. Furthermore, while \cite{LatentICASSP} was only applicable in fully observable \ac{ss} models, \name~presented here is designed to leverage its access to the state evolution model to track the state also in partially observable settings. This enables its application in various challenging scenarios, as also demonstrated in our numerical study reported in Section~\ref{sec:Empirical_Study}. 

Our design of \name~improves estimation performance by breaking the separation between feature extraction and filtering. In principle, one can claim that providing the prior $\hat{\gvec{x}}_{t|t-1}$ effectively delegates the filtering task to the instantaneous estimate \ac{dnn} for fully observable models, and renders the following filtering  step meaningless. However, our numerical findings reported in Section~\ref{sec:Empirical_Study} demonstrate that this is not the case, and that the system benefits from both prior-aided instants estimation as well as from the filtering operation based on that estimate.  Step 4 allows the resulting algorithm to operate without enforcing a Gaussian \ac{ss} model on the latent representation, as opposed to \cite{ zhou2020kfnet}. This is obtained as Algorithm~\ref{alg:LatKN} bypasses the need to model the stochasticity in the \ac{ss} model by using \acl{kn} \cite{revach2022kalmannet}. 
Unlike state-of-the-art \ac{dnn}-aided filters, such as the \acs{rkn} of \cite{becker2019recurrent}, we do not replace all the \ac{kf} procedures with \acp{dnn} and preserve the operation of the model-based filter. This allows us to systematically incorporate the available domain knowledge on the state evolution, improving performance and generalization, as demonstrated in Section~\ref{sec:Empirical_Study}. 

The hybrid model-based/data-driven design of \name~yields gains not only in accuracy and interpretability. It can also achieve faster inference speed compared to other model-based solutions or highly parameterised data driven models,  while supporting training with relatively limited  datasets, as demonstrated in Section~\ref{sec:Empirical_Study}. The computation complexity for each time step is linear in the dimensions of the trainable modules, being the complexity order of inferring using a \ac{dnn}, while its augmentation with the classic \ac{ekf} enables using relatively compact architectures, as we do in Section~\ref{sec:Empirical_Study}. Moreover, while \name~follows the operation of the \ac{ekf}, it does not involve Jacobian computations as in \eqref{eqn:Jacobians} or matrix inversion as in \eqref{eqn:kalman_gain_computaion} on each time step. This implies that \name~is a good candidate to apply for high dimensional \ac{ss} models and on computationally limited devices, compared to other model-based solutions, as well as data-driven approaches with a large volume of weights. Preserving the model-based operation of the \ac{kf} was shown to bring operational gains beyond accuracy and training complexity. For instance, it was shown in \cite{klein2022uncertainty} to enable extracting uncertainty on the estimates, and in \cite{revach2022unsupervised} to facilitate unsupervised learning. We leave the exploration of these properties for latent-space learned filtering for future work. 

%

\begin{figure}
\centering
\includegraphics[width=0.4\columnwidth]{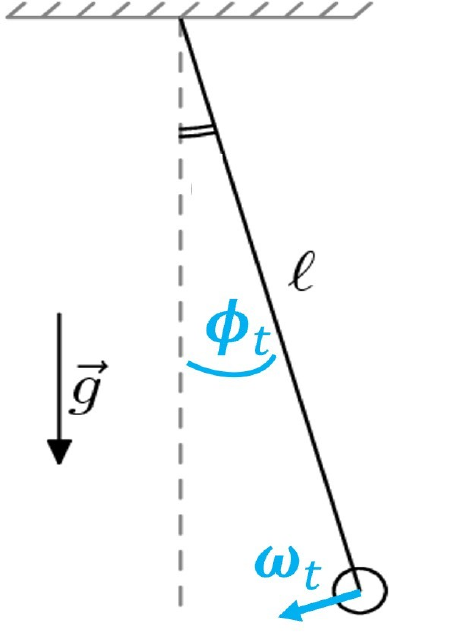}
\caption{Pendulum: physical setup and state variables.}
\vspace{0.2cm}
\label{fig:Pendulum: Physical setup}
\end{figure}

\section{Empirical Study}
\label{sec:Empirical_Study}
In this section, a comprehensive numerical analysis is performed\footnote{The source code and hyperparameters used are available  at \url{https://github.com/KalmanNet/Latent_KalmanNet_TSP.git}} on the proposed \name~to assess its performance. We consider two setups involving the tracking of a dynamic system from visual measurements: The first study detailed in Subsection~\ref{ssec:Design_Steps_Contribution} considers is a partially observable dynamic system representing the tracking of a pendulum. It is used to critically examine the design steps of \name~and to evaluate the contribution of each of its components.
The second study, presented in Subsection~\ref{ssec:Analyzing_Learned_KF_in_Latent_Space}, considers the Lorenz attractor chaotic system, which is an observable dynamic system. This setup is used to compare our proposed \name~against both model-based and \acl{dd} techniques across various scenarios, with both full and partial domain knowledge.  



  \begin{figure}
\includegraphics[width=\columnwidth]{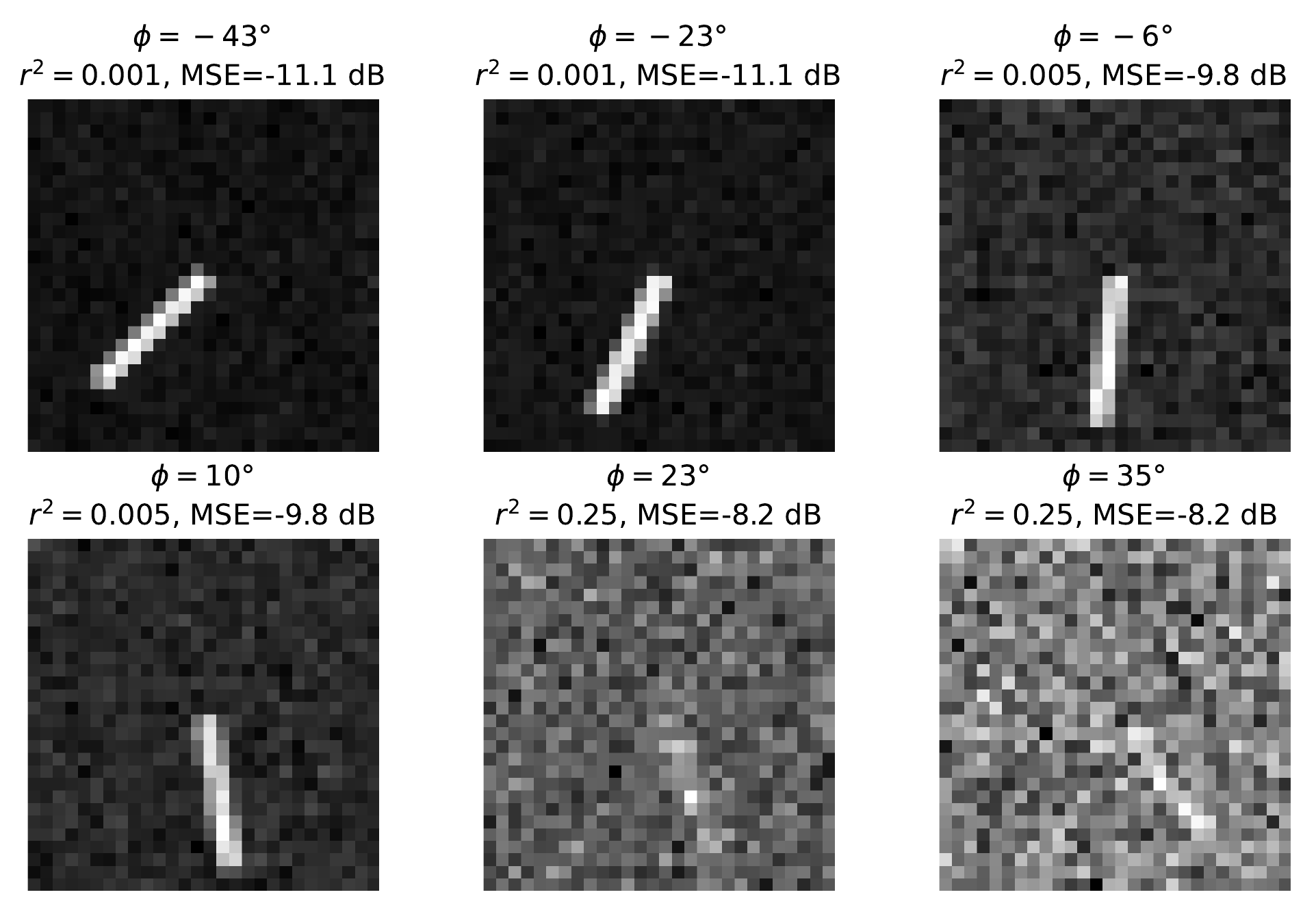}
    \caption{{Pendulum: several representative gray-scale observations along with their corresponding angle variable $\phi$, set to be the ground truth. In addition, the variance Gaussian noise level $r^2$ that was added to the image, and the \ac{mse} achieved by \name.}}
    \label{fig:Pendulum observations}
  \end{figure}

\begin{figure*} 
\begin{center}
\begin{subfigure}[pt]{0.99\columnwidth}
\includegraphics[width=1\columnwidth]{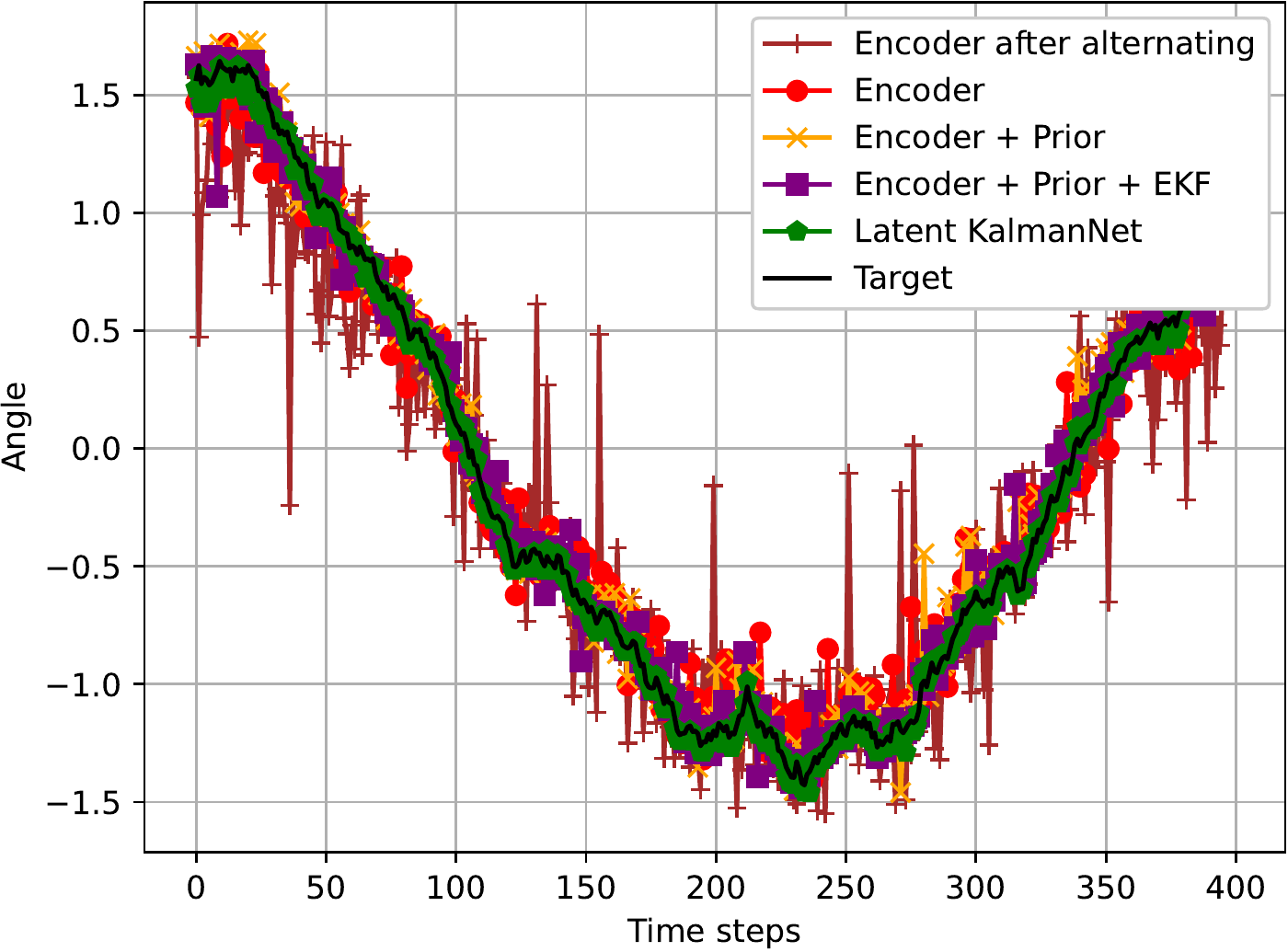}
\caption{Trajectory of $400$ time instances} 
\label{fig:trajecory estimation} 
\end{subfigure}
\begin{subfigure}[pt]{0.99\columnwidth}
\includegraphics[width=1\columnwidth]{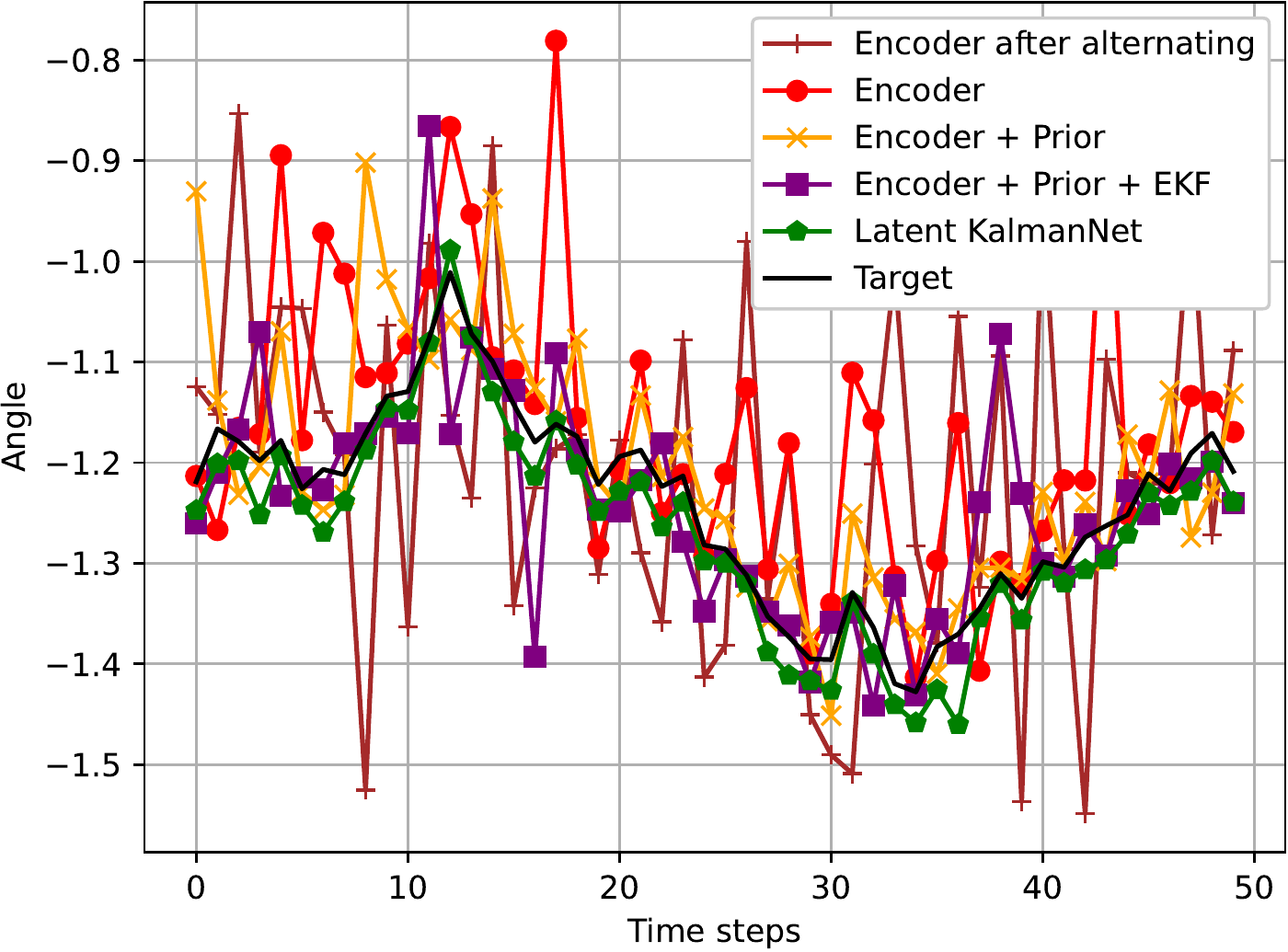}
\caption{Zoom in on $50$ time instances}
\label{fig:zoom}
\end{subfigure}
\vspace{0.5cm}
\caption{{Pendulum:} State estimation of the angle variable for a single trajectory realization}
\label{fig:Trajectory estimation pendulum} 
\end{center} 
\vspace{0.1cm}
\end{figure*}


\subsection{Pendulum Data}
\label{ssec:Design_Steps_Contribution}
We commence our numerical study by comprehensively evaluating the impact of each design step outlined in Section~\ref{sec:Learned_Kalman_Filtering_in_the_Latent_Space} over the Pendulum setting, further detail in the following.

\subsubsection{\ac{ss} Model}
In the considered \ac{ss} model,  the vector $\gvec{x}_{t}$ represents the state of an oscillating pendulum,   encompassing both the angle ${\phi}_{t}$ and the angular velocity $\omega_{t}$, i.e., $\gvec{x}_{t} = 
    [{\phi}_{t}, \omega_{t}]^\top$.
We focus on tracking the angle ${\phi}_{t}$ along the trajectory of a pendulum movement that is released from rest at a pre-defined point, i.e., the \ac{mse} is reported with respect to ${\phi}_{t}$. 
 The state evolution model of the pendulum is defined by mechanical system laws, making it highly nonlinear in nature, as given in the following equation 
\begin{equation}
\label{eqn:pendulum_dynamic_model}
    \gvec{x}_{t} = \begin{bmatrix}
1 & \Delta_{t} \\
0 & 1
\end{bmatrix} \cdot \gvec{x}_{t-1} - \frac{{g}}{\ell} \cdot \begin{bmatrix}
1/2 \cdot \Delta_{t}^2 \\
\Delta_{t}
\end{bmatrix} \cdot \sin({\phi}_{t-1}) + \gvec{e}_t.
\end{equation} 
In \eqref{eqn:pendulum_dynamic_model},  $\Delta_{t}$ denotes the sampling interval, dictating the time difference between consecutive observations, and $\gvec{e}_t$ is an i.i.d. zero-mean Gaussian noise with covariance $\gvec{Q}=q^2\cdot \gvec{I}$, where $q^2 = 0.1$. The gravitational acceleration is set to a constant value of $g=9.81 \sbrackets{\gscal{m}/\gscal{sec}^2}$, and the length of the string is represented by $\ell$. Fig.~\ref{fig:Pendulum: Physical setup} illustrates the physical pendulum setup.

The observations $\gvec{y}_t$ are $28\times 28$ gray-scale images generated from the sampled trajectories of the  pendulum. The images capture the pendulum's dynamic movements as if they were taken by a camera set in front of the system, corrupted by  i.i.d. Gaussian  observation noise $\gvec{v}_t$ with covariance $\gvec{R}=r^2\cdot \gvec{I}$, where $r^2\in{0.001,\ldots,0.25}$. Fig.~\ref{fig:Pendulum observations} shows several representative visual observations of a given trajectory, with different added noise variance $r^2$. 
As only the angle can be recovered from a single image, this setting represents a partially observable \ac{ss} model (see Subsection~\ref{ssec:Problem_Formulation}) with $\gvec{P}=[1,0]$. 
{We use this model to generate $D=1,000$ trajectories of length $T=200$ which comprise the data set $\mySet{D}$ as \eqref{eqn:DataSet}, with additional $100$ trajectories for evaluation.}

\subsubsection{Tracking Methods}
To this end, we evaluate the following approaches, representing the steps in designing \name:
\begin{itemize}
    \item  {\em Encoder} (Step 1):
We implement a purely \acl{dd}, model-agnostic convolutional encoder, comprised of 3 convolutional layers, followed by 2 \ac{fc} layers, and $p=1$ output neuron, as detailed in Table  \ref{table: Encoder Architecture}. The encoder is trained in a supervised manner on the dataset $\mySet{D}$ \eqref{eqn:DataSet} with loss function \eqref{eqn:loss_function_encoder} to map each $\gvec{y}_t$ into an estimate of the observable state variable  ${\phi}_{t}$. 

\begin{table}
\centering
\small{
\caption{Encoder Architecture}
\label{table: Encoder Architecture}
\centering
    \begin{tabular}{|c|c|c|c|c|}
    \rowcolor{lightgray}
         \hline
         Layer & Filter Size & Stride & Channels & Output size \\ [0.5ex]
         \hline
         \hline
         Input & - & - & - & 1x28x28 \\
         \hline
          conv2D & 3x3 & 2 & 8 & 8x14x14\\
         \hline
          ReLU & - & - & - & 8x14x14\\
         \hline
          Batch Norm & - & - & 8 & 8x14x14 \\
         \hline
          conv2D & 3x3 & 2 & 16 & 16x7x7 \\
         \hline
          ReLU & - & - & - & 16x7x7\\
         \hline
          Batch Norm & - & - & 16 & 16x7x7 \\
         \hline
          conv2D & 3x3 & 2 & 32 & 32x4x4\\
         \hline
          ReLU & - & - & - & 32x4x4\\
         \hline
          Batch Norm & - & - & 32 & 32x4x4\\
         \hline
          Flatten & - & - & - & 512\\
         \hline
          FC & - & - & 32 & 32\\
         \hline
          ReLU & - & - & - & 32\\
         \hline
          FC & - & - & $p$ & $p$\\
         \hline
    \end{tabular}}
\end{table}

\begin{figure}
\centering
\includegraphics[width=\columnwidth]{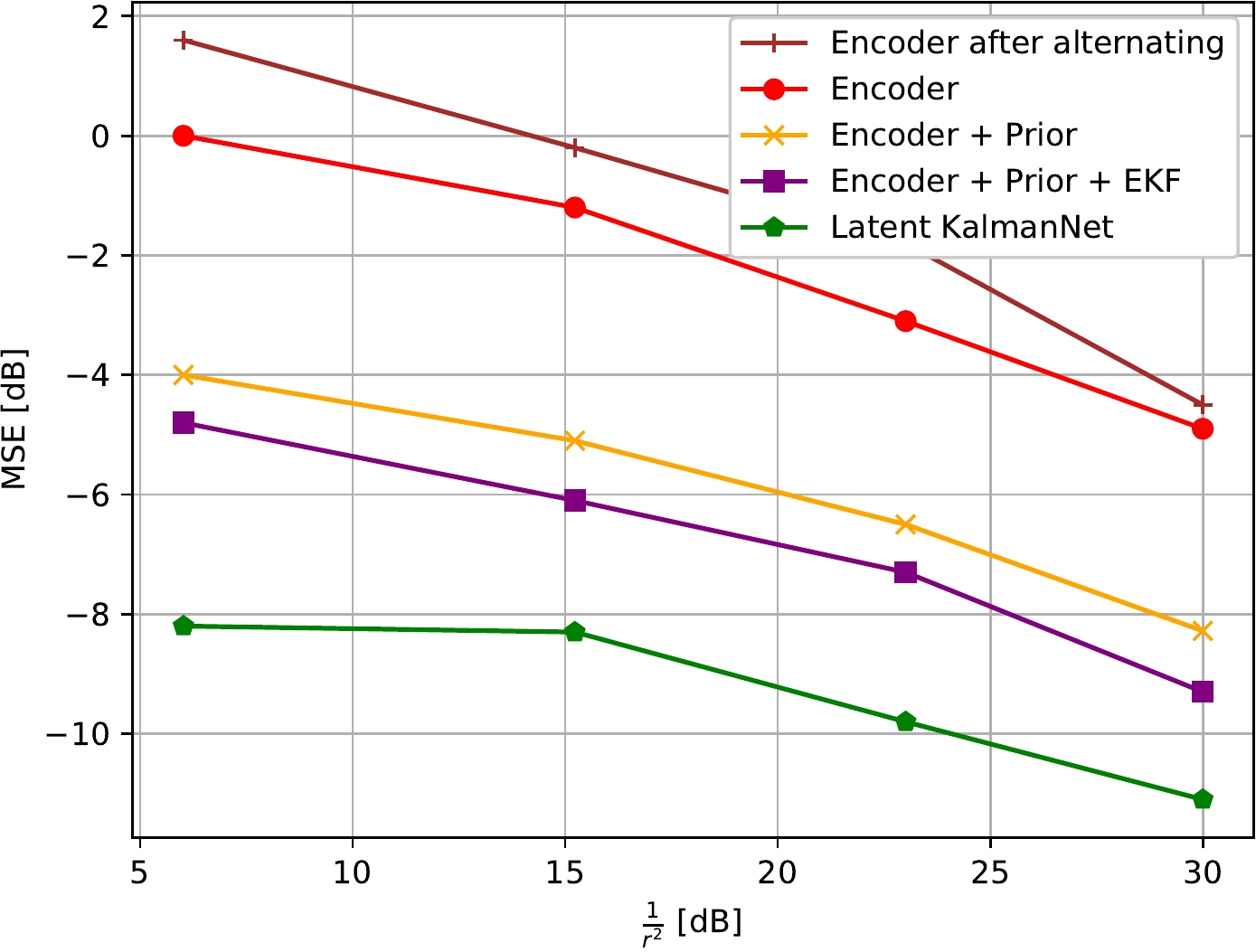}
\caption{{Pendulum: Design steps contribution - MSE vs. different Gaussian noise variance added to the images $\frac{1}{r^2}$.}} 
\label{fig:Design_steps_contribution}
\end{figure}

\begin{figure*}
  \begin{subfigure}[a]{0.5\columnwidth}
    \centering
    \includegraphics[width=1\columnwidth]{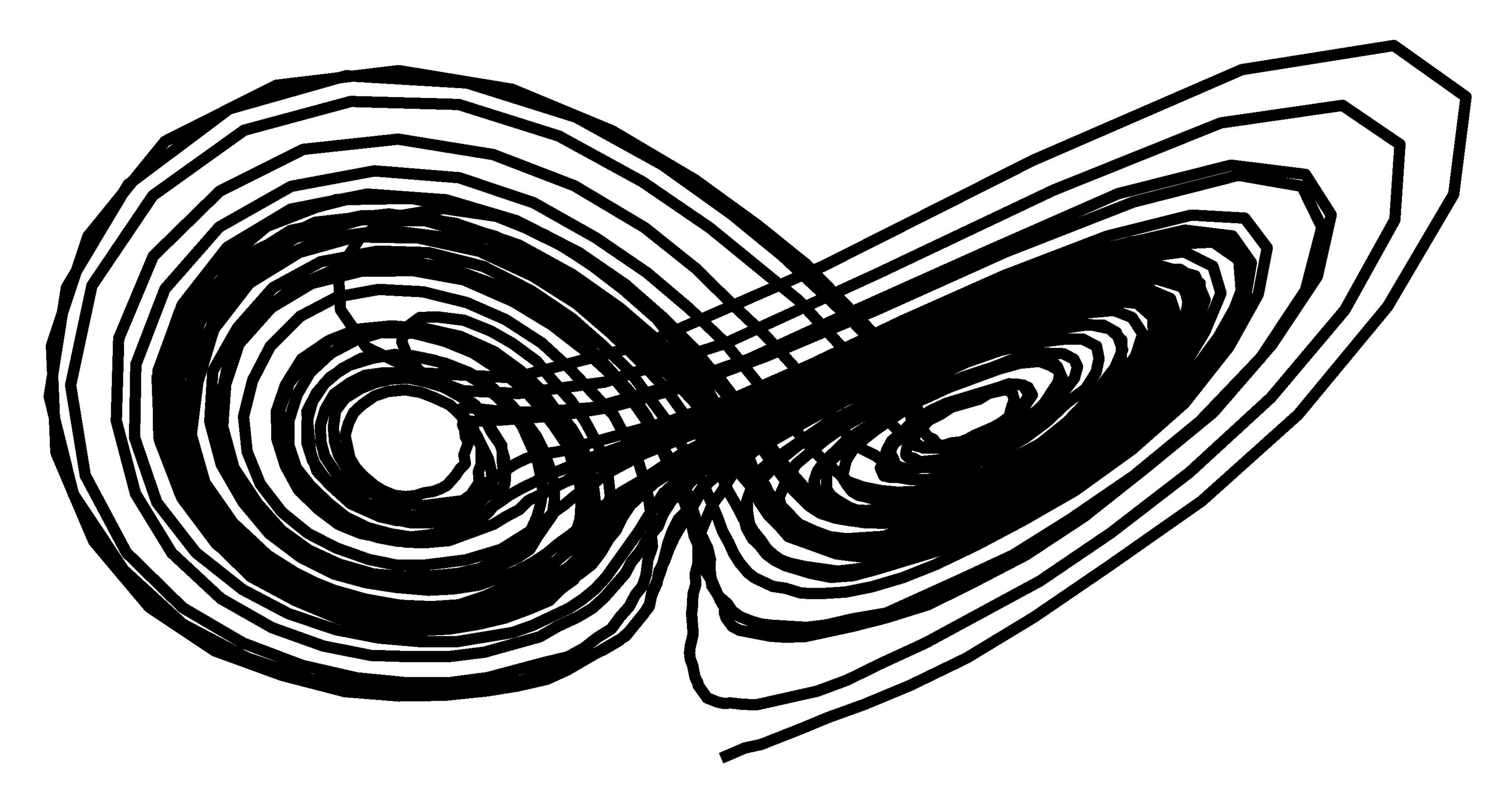}
    \caption{Ground truth.}
    \label{fig:figure1b}
  \end{subfigure}
  \hfill
  \begin{subfigure}[a]{0.5\columnwidth}
    \centering
    \includegraphics[width=1\columnwidth]{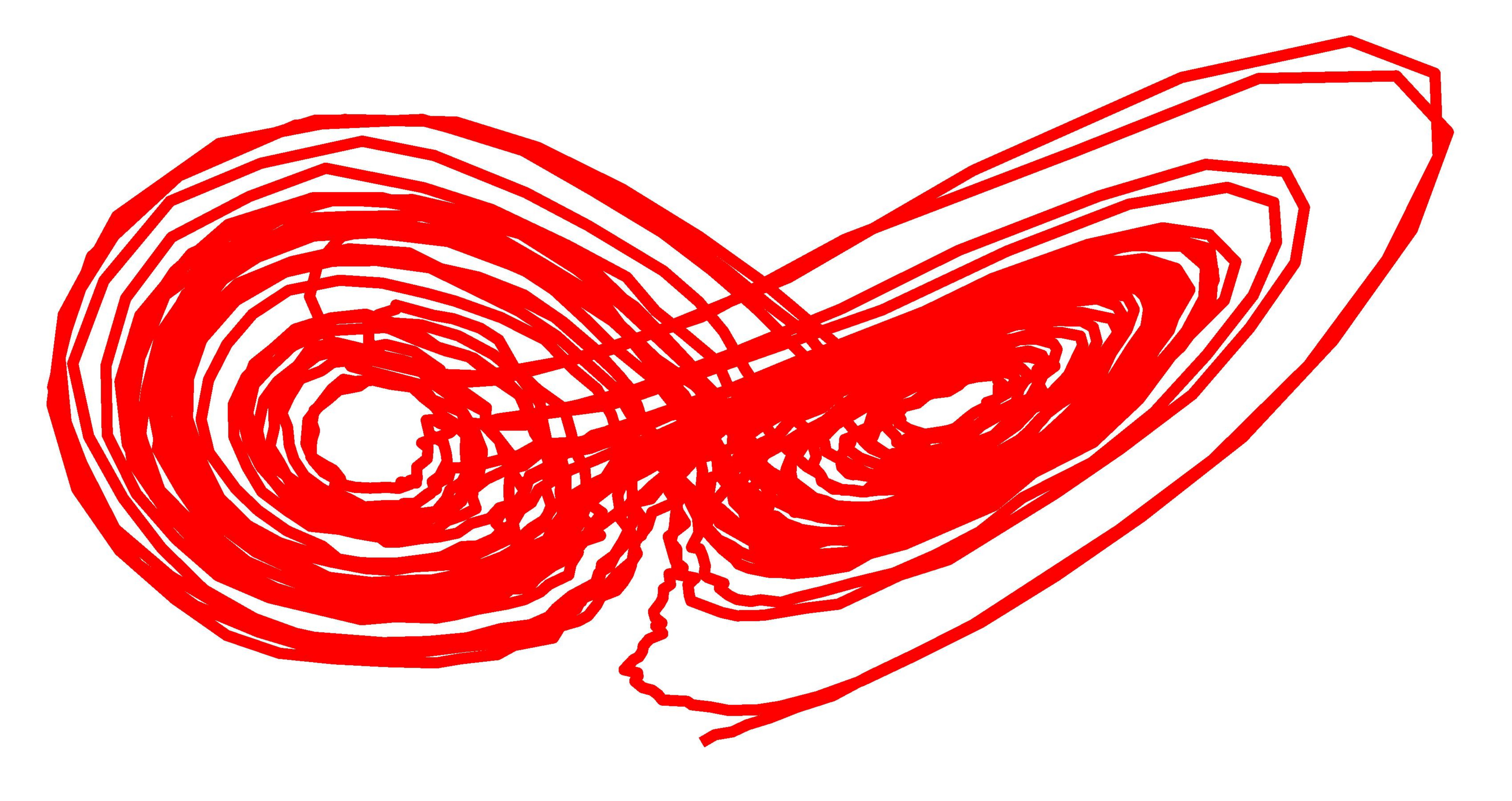}
    \caption{Encoder}
    \label{fig:figure2a}
  \end{subfigure}
  \hfill
  \begin{subfigure}[a]{0.5\columnwidth}
    \centering
    \includegraphics[width=1\columnwidth]{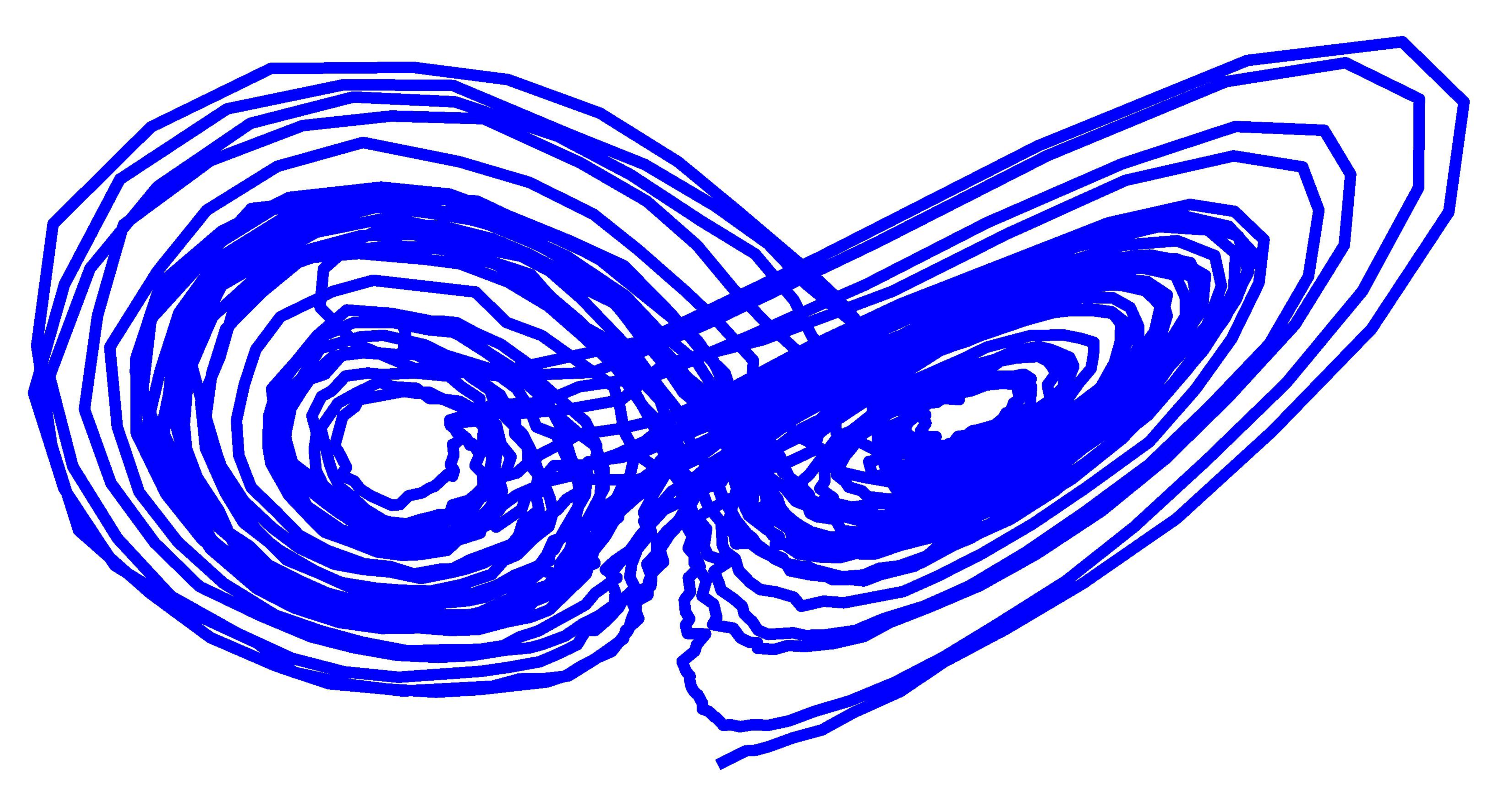}
    \caption{RKN}
    \label{fig:figure2b}
  \end{subfigure}
  \hfill
  \begin{subfigure}[a]{0.5\columnwidth}
    \centering
    \includegraphics[width=1\columnwidth]{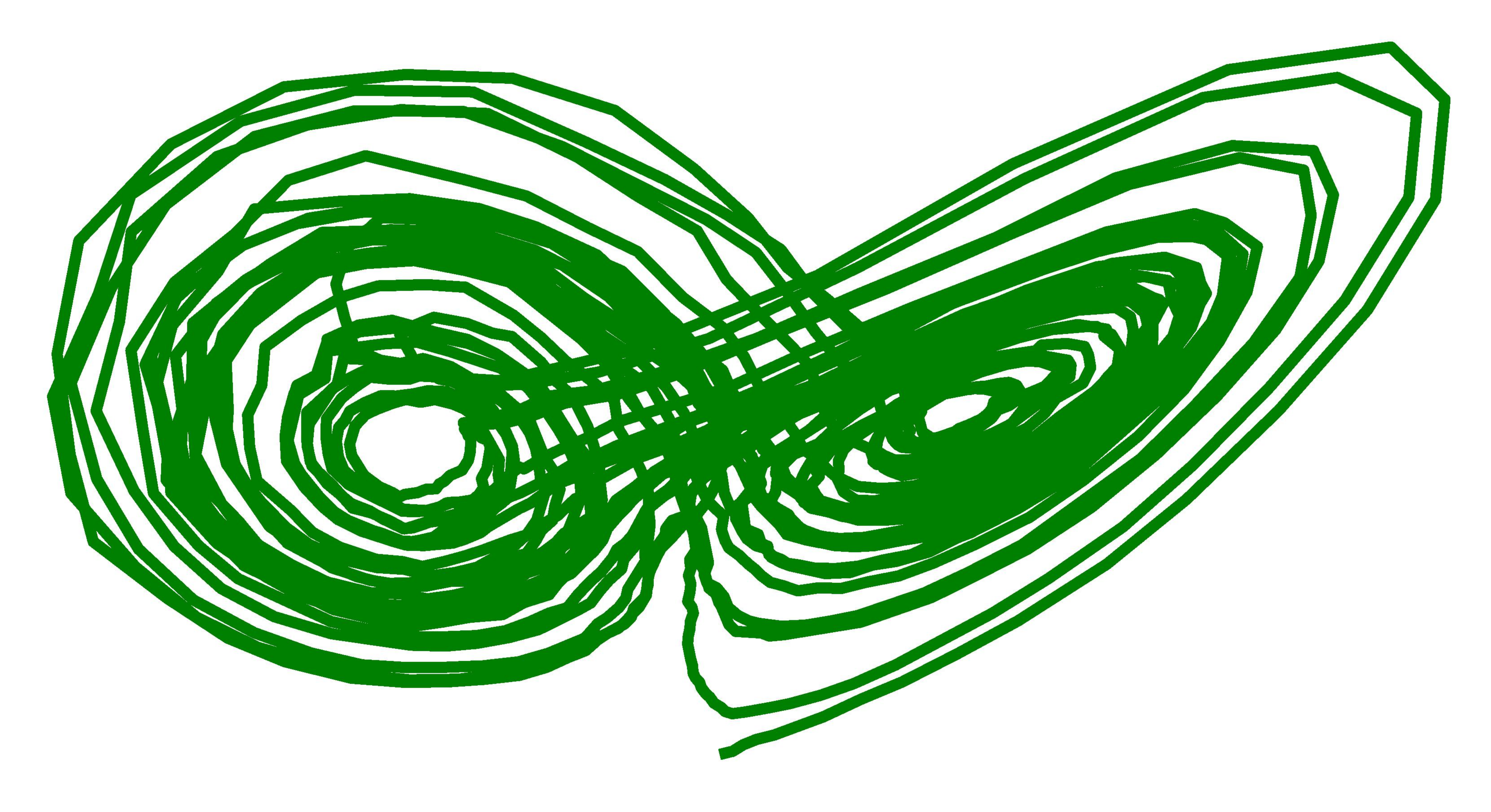}
    \caption{Latent-KalmanNet}
    \label{fig:figure2c}
  \end{subfigure}
  \vspace{0.3cm}
  \caption{Lorenz attractor: ground truth trajectory of the state vs. trajectories estimated with the different methods.}
  \label{fig:Lorenz attractor data illustration}
 \end{figure*}

\item   {\em Encoder + Prior} (Step 2):
We modify the encoder architecture, detailed in Table \ref{table: Encoder Architecture} by incorporating a prior, $\gvec{f}(\hat{\gvec{x}}_{t-1})$, as an additional input to the observed image $\gvec{y}_t$. The prior undergoes an \ac{fc} layer and concatenates to the flattened version of the extracted features, as illustrated in Fig. \ref{fig:Encoder with prior block diagram}. The output of this encoder still represents the estimate of ${\phi}_{t}$, but is fused with an estimate of unobservable angular velocity variable, obtained by applying the state evolution $\gvec{f}(\hat{\gvec{x}}_{t-1})$.

\item {\em Encoder + Prior + EKF} (Step 3):
On top of the trained encoder with prior of Step 2, we apply an \ac{ekf} with the observation function set to $\gvec{h}(\gvec{x})=\gvec{P}\gvec{x}$. The   variance of the state evolution noise is selected through grid search, while the observation noise is determined by the  empirical estimate loss at the output of the encoder. 

\item  {\em \name} (Step 4): 
The proposed \name~uses the Encoder + Prior of Step 2, and combines it with \acl{kn} based implemented using Architecture 2 of \cite{revach2022kalmannet}. \name~is trained using Algorithm~\ref{alg:Alternating}. 
\end{itemize}


\subsubsection{Results}
In Fig. \ref{fig:Design_steps_contribution} we compare the \ac{mse} averaged over  {$100$ test trajectories} achieved by the considered methods in recovering angle, i.e., the observable entry of the state variables. The findings in Fig.~\ref{fig:Design_steps_contribution} reveal the individual contribution of each of the design steps comprising \name. There, it is shown that including a prior based on the state evolution model notably improves the performance of an encoder in recovering the observable state variables. Moreover, using this estimate as latent features and employing \ac{ekf} tracking in latent space further improves performance, though less dramatically; However, replacing the \ac{ekf} with \acl{kn} that is jointly trained along with the latent representation as \name~achieves substantial improvements in \ac{mse}. We also depict in Fig.~ \ref{fig:Design_steps_contribution} the \ac{mse} computed at the output of the encoder of \name, where it is shown that the latent representation learned is not an accurate estimate of the state, being in fact worse than an instantaneous encoder. However, this  representation is learned such that it facilitates tracking in latent space, as evident by the superior performance of \name.

While Fig.~\ref{fig:Design_steps_contribution} reports the averaged \ac{mse}, in Fig.~\ref{fig:Trajectory estimation pendulum} the superiority of \name~is showcased when tracking a single trajectory. There, the improved tracking quality is highlighted by observing a trajectory spanning $400$ time instances in Fig. \ref{fig:trajecory estimation}, and we also zoom in on $50$ time instances in Fig.~\ref{fig:zoom} to improve visualization. These results demonstrate that \name's principled incorporation of the state evolution model knowledge while jointly learning the latent representation with the tracking algorithm results in improved angle estimation and smoother tracking. 



\subsection{Comparative Evaluation of \name}
\label{ssec:Analyzing_Learned_KF_in_Latent_Space}
Next, we present a comprehensive numerical evaluation of \name~and its performance in terms of \ac{mse} and latency. To that aim, we simulate various scenarios involving the nonlinear Lorenz attractor \ac{ss} model {detailed in the following.} 

  \begin{figure}
\includegraphics[width=\columnwidth]{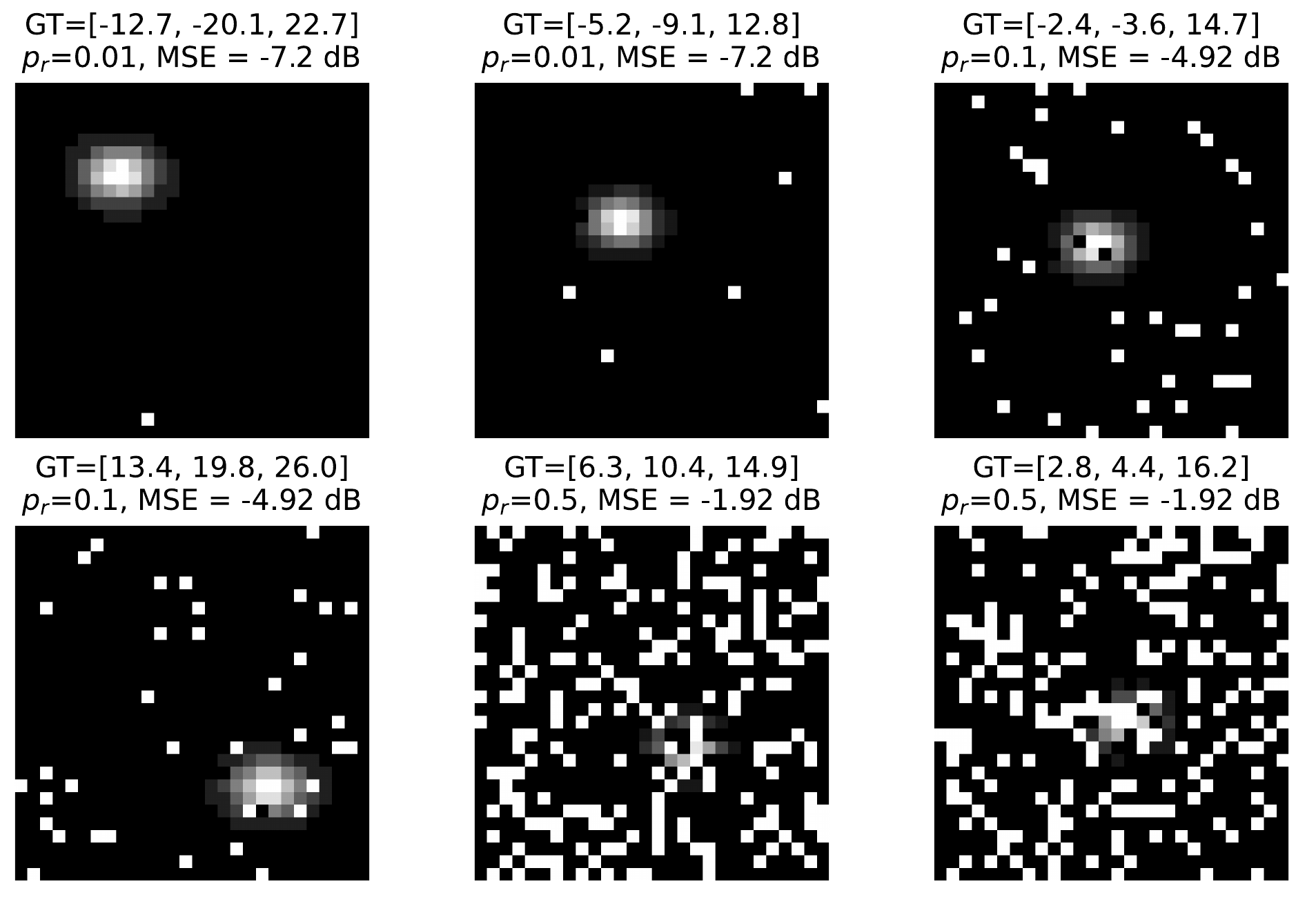}
    \caption{Lorenz attractor: several representative gray-scale observations along with their corresponding state value $\gvec{x}$, set to be the ground truth (GT). In addition, the \text{S\&P} noise probability that was added to the image, and the \ac{mse} achieve by \name.}
    \label{fig:Lorenz observations}
  \end{figure}

\subsubsection{\ac{ss} Model} 
Here, the state vector $\gvec{x}_{t}$ is a three-dimensional chaotic solution to the Lorenz system of ordinary differential equations. The system describes chaotic particle movement  sampled into discrete time intervals \cite{gilpin2021chaos}. The result is a nonlinear state evolution model of the continuous-time process, showcasing the dynamic interplay of forces shaping the chaotic particle's movement. The noise-free state-evolution equation is obtained from the differential equation
\begin{equation}
\label{eqn:Lorenz_dynamic_model}
\frac{d\gvec{x}_t}{dt}  = \mathbf{A}(\gvec{x}_t) \cdot \gvec{x}_t; \quad \mathbf{A}(\gvec{x}_t) = \begin{bmatrix}
-10 & 10 & 0 \\
28 & -1 & -x_1 \\
0 & x_1 & -8/3 
\end{bmatrix}
\end{equation}

\begin{figure*} 
\begin{center}
\begin{subfigure}[pt]{0.99\columnwidth}
\includegraphics[width=1\columnwidth]{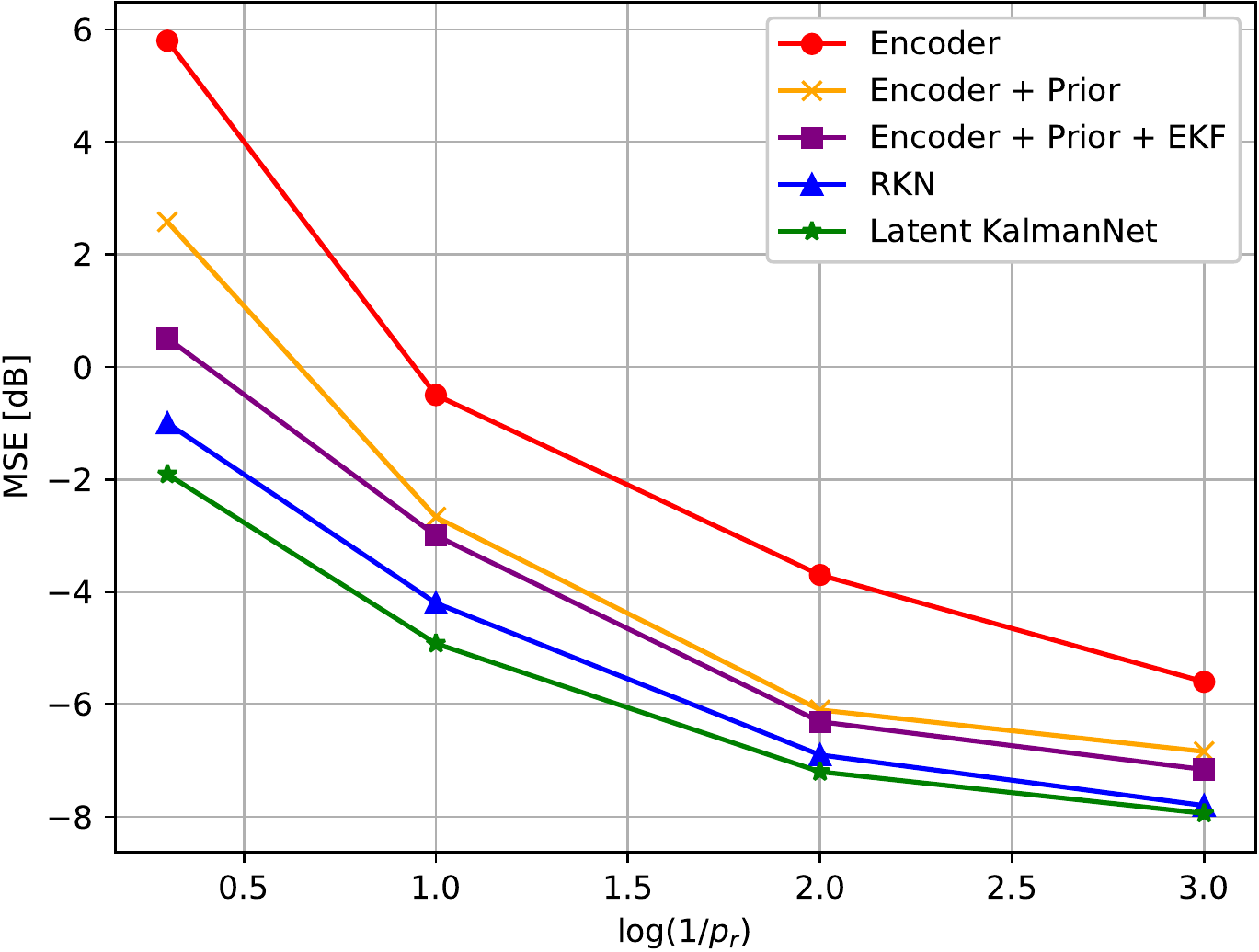}
\caption{Same trajectories length: $T_{\rm train} = T_{\rm test}=200$.} 
\label{fig:baseline} 
\end{subfigure}
\begin{subfigure}[pt]{0.99\columnwidth}
\includegraphics[width=1\columnwidth]{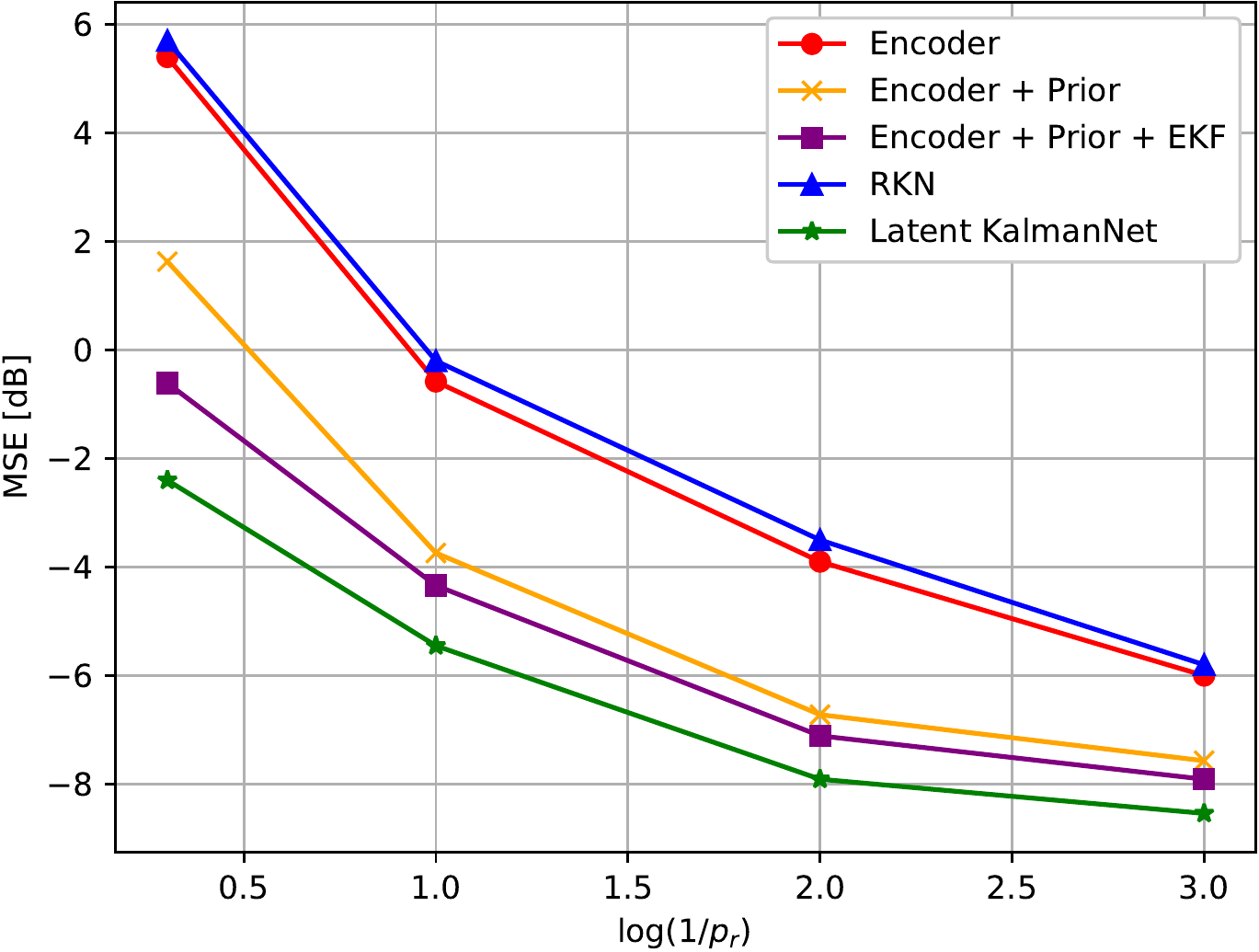}
\caption{Different trajectories length: $T_{\rm train} = 200, T_{\rm test}=2000$.}
\label{fig:long}
\end{subfigure}
\vspace{0.3cm}
\caption{Lorenz Attractor: Performance with full domain knowledge. MSE vs. observations \text{S\&P} noise level.}
\end{center} 
\vspace{-0.3cm}
\end{figure*}

The model is converted into a discrete-time state-evolution model by repeating the
steps used in \cite{garcia2019combining}. First, we sample the noiseless process with sampling interval $\Delta_{t}$ and assume that  can be kept constant in a small neighborhood of $\gvec{x}_t$ ; i.e.,
\begin{equation}
\mathbf{A}(\gvec{x}_t) \approx \mathbf{A}(\gvec{x}_{t+\Delta _t}).
\end{equation}
Then, the continuous-time solution of the differential system
\eqref{eqn:Lorenz_dynamic_model}, which is valid in the neighborhood of $\gvec{x}_t$ for a short time
interval $\Delta_{t}$, is
\begin{equation}
\label{eqn:Lorenz_dynamic_model_exp}
\gvec{x}_{t+\Delta _t} = \exp(\mathbf{A}(\gvec{x_t}) \cdot \Delta t) \cdot \gvec{x}_t.
\end{equation}
Finally, we take the Taylor series expansion of \eqref{eqn:Lorenz_dynamic_model_exp} and a finite
series approximation (with $J$ coefficients), which results 
\begin{equation}
\label{eqn:Lorenz_dynamic_model_final}
\mathbf{F}(\gvec{x}_t)\triangleq \exp(\mathbf{A}(\gvec{x}_t) \cdot \Delta t) = \mathbf{I} + \sum_{j=1}^{J} \frac{(\mathbf{A}(\gvec{x}_t) \cdot \Delta_t)^j}{j!}
\end{equation}
The resulting discrete-time evolution process is given by
\begin{equation}
\label{eqn:Lorenz_dynamic_model_approx}
\gvec{x}_{t+1} = \gvec{f}(\gvec{x}_t) = \gvec{F}(\gvec{x}_t) \cdot \gvec{x}_t.
\end{equation}
The discrete-time state-evolution model presented in \eqref{eqn:Lorenz_dynamic_model_approx}, is augmented with additional {zero-mean Gaussian noise with i.i.d. entries of variance $q^2=0.005$}, obtaining a noisy state-evolution representation as in \eqref{eqn:SSModelState}. An illustration of the state trajectory is given in the left side of Fig.~\ref{fig:Lorenz attractor data illustration}.

%
%
\begin{table}
\begin{center}
{
\begin{tabular}{|c|c|c|c|c|}
\hline
\rowcolor{lightgray}
Noise level $-\log(p_r)$ & 0.3 & 1 & 2 & 3 \\ [0.5ex]
\hline \hline
Encoder & $5.8$ & $-0.5$ & $-3.7$ & $-5.6$ \\
        & $\pm{1.1}$ & $\pm{1.3}$ & $\pm{0.85}$ & $\pm{0.9}$ \\
\hline
Encoder+Prior & $2.58$ & $-2.67$ & $-6.1$ & $-6.84$ \\
              & $\pm{0.5}$ & $\pm{0.58}$ & $\pm{0.48}$ & $\pm{0.52}$ \\
\hline
Encoder+Prior+EKF & $0.51$ & $-3$ & $-6.31$ & $-7.16$\\
                  & $\pm{0.35}$ & $\pm{0.41}$ & $\pm{0.38}$ & $\pm{0.32}$\\
\hline
RKN & $-1$ & $-4.2$ & $-6.9$ & $-7.8$ \\
    & $\pm{2.1}$ & $\pm{1.5}$ & $\pm{1.3}$ & $\pm{1.1}$\\
\hline
\name & \textbf{-1.91} & \textbf{-4.92} & \textbf{-7.2} & \textbf{-7.94}\\
      & $\pm{\textbf{0.06}}$ & $\pm{\textbf{0.1}}$ & $\pm{\textbf{0.07}}$ & $\pm\textbf{0.12}$\\
\hline
\end{tabular}}
\end{center}
\caption{Lorenz Attractor: Numeric \ac{mse} values for the setting reported in Fig~\ref{fig:baseline}, including standard deviation in the \ac{mse}.}
\label{tbl:baseline}
\end{table}

We emulate visual representation of the movement described by $\gvec{x}_t$ in the form of $28 \times 28$ matrices. To represent visual observations used in particle tracking, e.g., \cite{bayle2021single}, the sensing function evaluated at coordinate $\gvec{c}\in\mathbb{R}^2$ for state value $\gvec{x}=[x_1, x_2, x_3]^\top$ is modeled a Gaussian point spread function whose intensity depends on the lateral state coordinate, where we use
\begin{equation}
\label{eqn:h_function}
\gvec{h}\brackets{\gvec{c};\gvec{x}} \!=\!
10 \exp\brackets{
\frac{-1}{{2 x_3}}{\norm{\gvec{c}-\begin{bmatrix} x_1 \\ x_2 \end{bmatrix}}^2}}.
\end{equation}
The observations are corrupted by  Salt and Pepper \text{(S\&P)} noise, modeled as an i.i.d. scaled Bernoulli vector with probability $p_r$. This type of noise is common in digital images and can be caused by sharp and sudden disturbances in the image signal when transmitting images over noisy digital channels.
  Representative visual observations of a given trajectory are depicted in Fig.~\ref{fig:Lorenz observations}. All considered tracking algorithms have access to the same dataset \eqref{eqn:DataSet}. Unless otherwise stated, the data was generated from the Lorenz attractor \ac{ss} model with Taylor order of $J= 5$, sampling interval of $\Delta _t = 0.02$, and a trajectory length of $T = 200$. 

\begin{figure*}
\begin{center}
\begin{subfigure}[pt]{0.99\columnwidth}
\includegraphics[width=1\columnwidth]{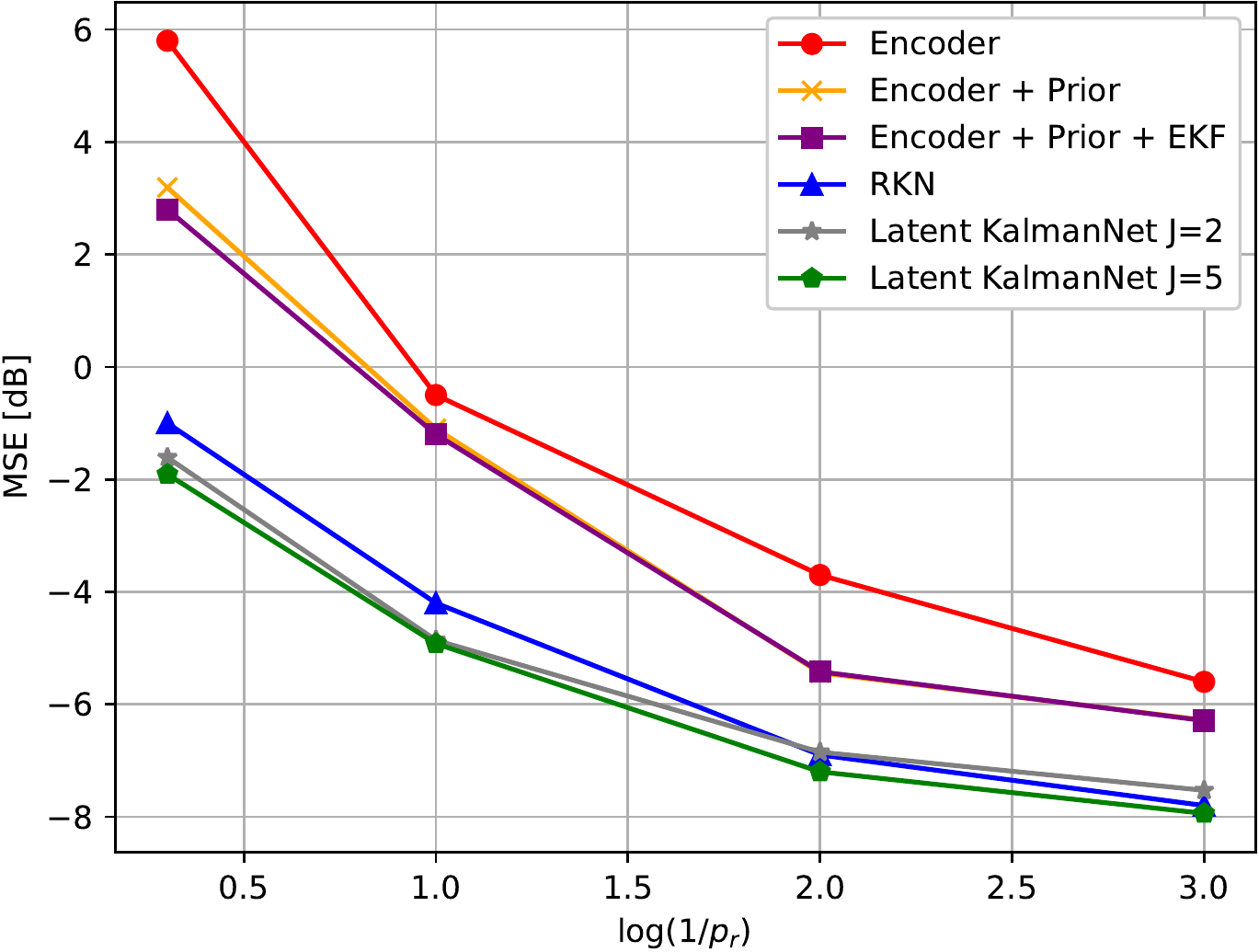}
\caption{Mismatch due to state-evolution Taylor expansion.}
\label{fig:Miss-match_state_evolution_function.}
\end{subfigure}
\begin{subfigure}[pt]{0.99\columnwidth}
\includegraphics[width=1\columnwidth]{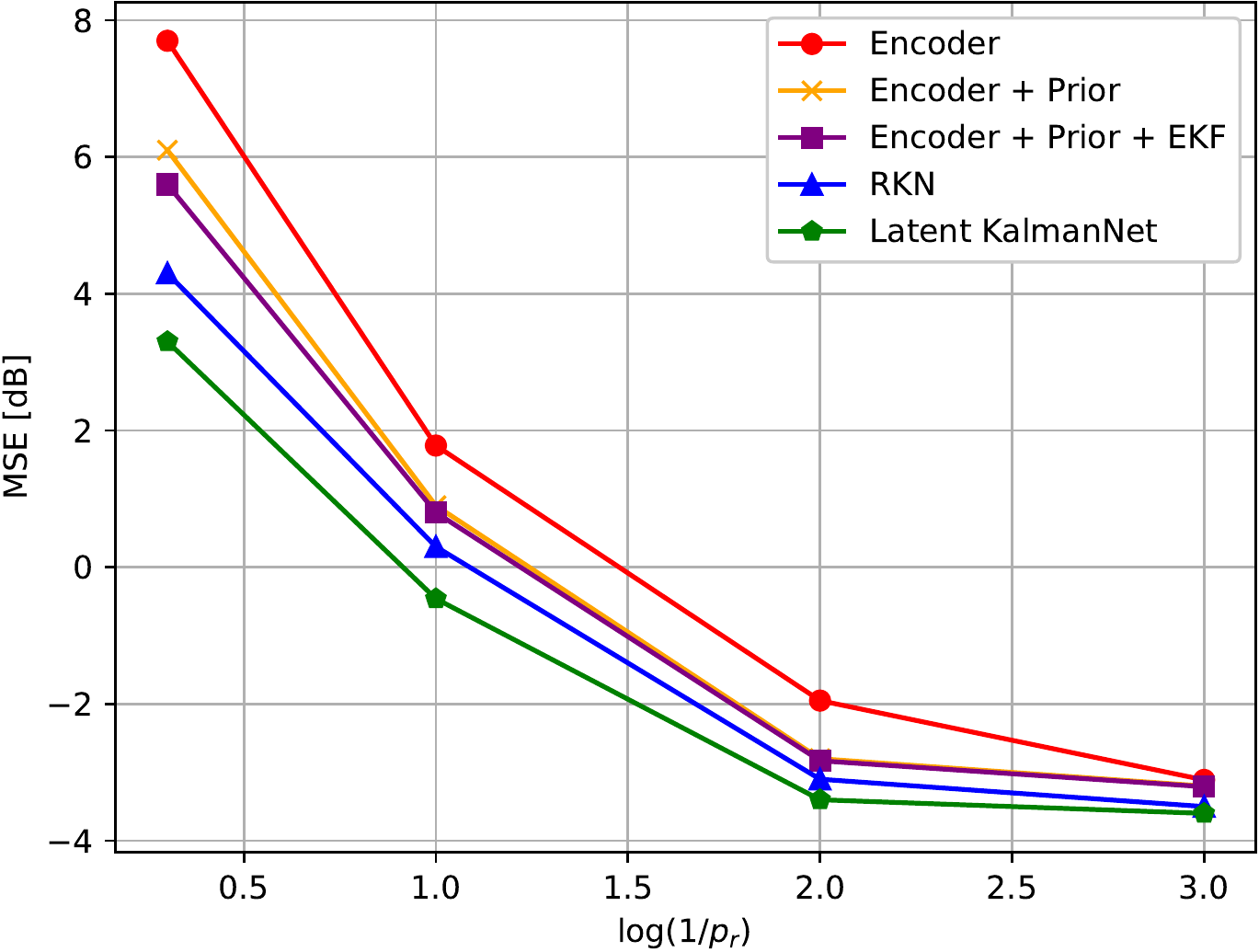}
\caption{Mismatch due to coarse sampling.}
\label{fig:Decimation.}
\end{subfigure}
\vspace{0.3cm}
\caption{Lorenz attractor: Performance with partially domain knowledge. MSE vs. observations \text{S\&P} noise level.}
\end{center}
\vspace{-0.3cm}
\end{figure*}

\subsubsection{Tracking Methods}
The following experiments aim to assess the efficacy of \name~in comparison to benchmark \acl{dd} algorithms as well as with model-based tracking in latent space. In particular, we evaluate the following tracking algorithms:
\begin{enumerate}
    \item {\em Encoder:}  A  \acl{dd} convolutional encoder as in Table~\ref{table: Encoder Architecture}, i.e.,  comprised of three convolutional layers and two \ac{fc} layers with $p=3$ output neurons.
    \item  {\em Encoder + Prior:} The convolutional encoder with a prior estimate stacked to features provided to the first \ac{fc} layer.
    \item {\em Encoder + Prior + EKF:} 
    Model-based tracking in latent space by applying an \ac{ekf} with the observation function set to the identity matrix, i.e., $\gvec{h}(\gvec{x})=\gvec{x}$. The variance of the state evolution noise is selected through a greed search, while the observation noise is determined by the empirical \ac{mse} at the output of the encoder. 
    \item  {\em RKN:}  The \ac{rkn} of \cite{becker2019recurrent}, which is a leading \acl{dd} tracking algorithm that utilizes a high-dimensional factorized latent state representation.
    \item {\em \name:} The proposed \name~implemented using the above Encoder Prior along with \acl{kn} based on Architecture 2 of \cite{revach2022kalmannet}.  
\end{enumerate}

\subsubsection{Results}
 To assess \name~in terms of tracking performance, latency, and robustness, we evaluate \ac{mse}  as well as latency and complexity. 
As the S\&P noise is characterized by the probability $p_r$, when evaluating how performance scales with the noise level we report the \ac{mse} values versus $\log \frac{1}{p_r}$ for ease of visualization, where $p_r \in \{0.5,\ldots, 0.001\}$ is mapped into $\log \frac{1}{p_r} \in \{0.3,\ldots,3\}$.
We consider both the case of {\em full information}, where the \ac{ss} model parameter (e.g., the state evolution function $\gvec{f}(\cdot)$) is the same as that used for generating the trajectories, and the case of {\em partial information}, where this domain knowledge is mismatched. 

\textbf{Full Information:} 
Here, we  compare \name~to the benchmark algorithms, where all algorithms have access to the state-evolution function $\gvec{f}(\cdot)$  used during data generation.  

In our first experiment, the trajectory length presented during training was the same as in the test set $T_{\rm train} = T_{\rm test} = 200$. The resulting \acp{mse}, reported in Fig.~\ref{fig:baseline}, demonstrate that the proposed \name~achieves the lowest \ac{mse} for all considered noise levels. The improvement due to adding EKF on top of the pre-trained encoder, tracking in the new learned latent space, is much less notable here compared with the improvement in the Pendulum setup noted in Fig.~\ref{fig:Design_steps_contribution}. This follows since the S\&P noise yields a latent representation in which the distribution of the distortion cannot be faithfully approximated as being Gaussian, while the \ac{ekf}, as opposed to \acl{kn}, is designed for Gaussian \ac{ss} models. The purely \acl{dd} \acs{rkn} improves upon the model-based EKF, and is only slightly outperformed by the proposed \name. 

The superiority of \name~is also evident in Table \ref{tbl:baseline}, which reports the \acp{mse} along with their standard deviation, representing the confident intervals of the estimators. It is observed that the improved estimates of \name~are also consistently achieved more confidently, i.e., with smaller standard deviation, compared with  the competitor RKN. This behavior is also illustrated when observing a single filter trajectory in Fig.~\ref{fig:Lorenz attractor data illustration}, where the original state tracked is the one depicted on the left side, and the different models predictions are on the right.

Next, we examine the generalization of the filters to different trajectory lengths. This is done by training the systems with trajectories of length $T_{\rm train} = 200$ while  testing with notably longer trajectories of length $T_{\rm test}=2000$. Fig.~\ref{fig:long} demonstrates how the purely \acl{dd}  \acs{rkn} struggles to generalize, achieving performance that is similar to the stand-alone encoder. However our \name~successfully generalizes to a much longer trajectory, as it learns how to track based on both data, latent features, and domain knowledge, allowing it to cope with the \ac{ss} model and not overfit to the trajectory length.

\textbf{Partial Information:}
To evaluate the performance of \name~under partial model information, we consider two sources of model mismatch in the Lorenz attractor setup. First, we examine state-evolution mismatch due to the use of a Taylor series approximation of insufficient order. In this study, both \name~and the benchmark algorithms (\textit{Encoder + Prior} and \textit{Encoder + Prior + EKF}) use a crude approximation of the evolution dynamics obtained by computing \eqref{eqn:Lorenz_dynamic_model_final} with $J = 2$, while the data was generated with an order $J = 5$ Taylor series expansion. We set 
the trajectory length to $T = 200$ (both $T_{\rm train}$ and $T_{\rm test}$), and  $\Delta _t = 0.02$. The results, depicted in Fig.~\ref{fig:Miss-match_state_evolution_function.}, demonstrate that applying a model-based \ac{ekf} achieves performance, which  coincides with that of the \textit{Encoder + Prior}. This stems from the fact that mismatched model resulted in the \ac{ekf} being unable to incorporate the state evolution to improve tracking, and the grid search for identifying the most suitable noise  variance lead to the Kalman Gain computation such that the estimation relies almost solely on the instantaneous  observation. More interestingly, \name~with partial knowledge (of $J = 2$) learns to overcome this model mismatch and manages to come within a small gap with the performance of \name~with full knowledge (of $J = 5$), outperforming its benchmark counterparts operating with the same level of partial information and the \acl{dd} \ac{rkn}. These findings suggest that \name~is robust and effective, even when operating under partial model information of the system dynamics.

Next, we evaluate the performance of \name~in the presence of sampling mismatch. We generate data from the Lorenz attractor \ac{ss} model using a dense sampling rate $\Delta _t = 0.001$, and then sub-sample the corresponding observations by a ratio of $\frac{1}{20}$ to obtain a decimated process with sample spacing of $\Delta _t = 0.02$. This results in a mismatch between the \ac{ss} model and the discrete-time sequence, as the nonlinearity of the \ac{ss} model results in a difference in distribution between the decimated data and data generated directly with sampling interval $\Delta _t = 0.02$. Such scenarios correspond to the practical setting of mismatches due to processing of continuous-time signals using discrete-time approximations. 
For this setting we use the identity mapping for the sensing function $\gvec{h}(\cdot)$ and set $T = 200$ (both $T_{\rm train}$ and $T_{\rm test}$). The results, shown in Fig.~\ref{fig:Decimation.}, demonstrate that \name~outperforms both model-based tracking and the \acl{dd} \ac{rkn} as its combination of learning capabilities, along with the available model dynamic knowledge, allows it overcomes the mismatch induced by representing a continuous-time \ac{ss} model in discrete-time. As in the case with mismatched state evolution, we again observe that the model mismatches result in the \ac{ekf} being unable to improve performance, and achieving the same performance as that of instantaneous detection using an Encoder + Prior. 

%
%
\begin{table}[b]
\caption{Latency and complexity comparison}
\begin{center}
\small{
\begin{tabular}{|c|c|c|c|}
\hline
\rowcolor{lightgray}
& Encoder+EKF & RKN & \name \\ [0.5ex]
\hline \hline
Complexity (FP) & \textbf{15573 +
243} & 42426 & 15573 +
2712\\
\hline
Latency (sec) & 0.36 & 0.39 & \textbf{0.09} \\
\hline
\end{tabular}} 
\end{center}
\label{table:latency}
\end{table}
\vspace{0.1cm}

\textbf{Complexity and Latency:}
We conclude our numerical study by demonstrating that the performance benefits of \name~do not come at the cost of increased computational complexity and latency, as is often the case when using deep models. In fact, we show that it can achieve faster inference compared with both model-based \ac{ekf} in latent space and the \acl{dd} \ac{rkn}. To that aim, we provide an analysis of the average inference time of these tracking algorithms computed over a  test set comprised of $100$ trajectories, with $T_{\rm test}=200$ time steps each. Inference time is computed on the same platform for all methods, which is an 11th Gen Lenovo laptop with Intel Core i7, 2.80 GHz processor, 16 GB of RAM, and Windows 11 operating system. We also report the number of floating point operations required by each method, where the number of operations in applying a \ac{dnn} is given by its number of trainable parameters. 

The resulting complexity and latency measures of \name~compared with \textit{Encoder + Prior + EKF} and \textit{RKN} are reported in Table~\ref{table:latency}. These results reveal that \name~achieves the fastest inference time, being not only notably faster than the purely \ac{dnn}-based \ac{rkn}, but also from the model-based \ac{ekf}. The latter stems from the fact that the \ac{ekf} needs to compute Jacobians and matrix inversions on each time instance to produce its Kalman gain, which turns out to be slower compared with applying the compact \ac{rnn} used by \acl{kn} for the same purpose, while being amenable to parallelization and acceleration. The compactness of the internal \ac{rnn} of \name~results in it having a similar computational complexity compared with applying the \ac{ekf} in latent space. 
These results, combined with the performance and robustness gains of \name~noted in the previous studies, showcase the potential of \name~in leveraging both data and domain knowledge for tracking with high-dimensional data while coping with the challenging \ref{itm:Dist}-\ref{itm:unknown_h}.


\section{Conclusions}
\label{sec:conclusions}
In this work, we proposed a method for tracking based on complex observations with unknown noise statistics. 
Our proposed \name~combines \ac{dnn}-aided encoding with learned Kalman filtering based on  \acl{kn}  in the latent space, and designs these modules to mutually benefit one another in a synergistic manner. 
The training scheme of \name~exploits its interpretable  architecture to formulate alternate training between the two learnable components in order to learn a surrogate latent representation, which most facilitates tracking. Our empirical evaluations demonstrate that the proposed \name~successfully tracks from high-dimensional observations and generalizes to trajectories of different lengths. It also succeeds in working with partial domain knowledge of the state evolution function or sampling mismatches, and is shown to infer with low latency. 
%
%
\bibliographystyle{IEEEtran}
\bibliography{IEEEabrv,refs}
%
%
\appendix
%
%
\subsection{Additional Numerical Results}
The results presented in these tables provide a numerical description of previous experiments that were demonstrated visually in \ref{sec:Empirical_Study}. The tables offer a concise and precise way of presenting the MSE for different noise levels, allowing for reconstructing the findings. Each Table point the corresponded figure. The Tables conclude also confidence intervals (std), providing a measure of the precision of the estimate. In all tables our \name~has narrow interval, proving our empirical superiority and strengthen the reliability of our results. 
\vspace{0.1cm}
%
%
\begin{table}[htbp!]
\caption{Pendulum: Design steps contribution, Fig \ref{fig:Design_steps_contribution}}
\label{tbl:Design steps}
\begin{center}
\scriptsize{
\begin{tabular}{|c|c|c|c|c|}
\hline
\rowcolor{lightgray}
Noise level $-\log(r^2)$ & 6 & 15.2 & 23 & 30 \\ [0.5ex]
\hline \hline
Encoder after alternating & $1.6$ & $-0.2$ & $-1.8$ & $-4.5$ \\
        & $\pm{0.7}$ & $\pm{0.66}$ & $\pm{0.61}$ & $\pm{0.56}$ \\
\hline
Encoder & $0$ & $-1.2$ & $-3.1$ & $-4.9$ \\
              & $\pm{0.51}$ & $\pm{0.41}$ & $\pm{0.48}$ & $\pm{0.43}$ \\
\hline
Encoder+Prior & $-4$ & $-5.1$ & $-6.5$ & $-8.28$ \\
              & $\pm{0.37}$ & $\pm{0.32}$ & $\pm{0.38}$ & $\pm{0.34}$ \\
\hline
Encoder+Prior+EKF & $-4.8$ & $-6.1$ & $-7.3$ & $-9.3$\\
                  & $\pm{0.22}$ & $\pm{0.34}$ & $\pm{0.25}$ & $\pm{0.21}$\\
\hline
\rowcolor{cyan}
\name & ${-8.2}$ & ${-8.3}$ & ${-9.8}$ & ${-11.1}$\\
\rowcolor{cyan}
      & $\pm{0.18}$ & $\pm{0.2}$ & $\pm{0.14}$ & $\pm{0.11}$\\
\hline
\end{tabular}}
\end{center}
\end{table}

%
%
\begin{table}[htbp!]
\label{tbl:long}
\caption{Lorenz: Performance with full domain knowledge. Different trajectories length, Fig \ref{fig:long}}
\begin{center}
\scriptsize{
\begin{tabular}{|c|c|c|c|c|}
\hline
\rowcolor{lightgray}
Noise level $-\log(p_r)$ & 0.3 & 1 & 2 & 3 \\ [0.5ex]
\hline \hline
RKN & $5.7$ & $-0.2$ & $-3.5$ & $-5.8$ \\
        & $\pm{1.6}$ & $\pm{2.4}$ & $\pm{2.8}$ & $\pm{1.5}$ \\
\hline
Encoder & $5.4$ & $-0.58$ & $-3.9$ & $-6$ \\
              & $\pm{0.79}$ & $\pm{0.71}$ & $\pm{0.75}$ & $\pm{0.8}$ \\
\hline
Encoder+Prior & $1.63$ & $-3.74$ & $-6.72$ & $-7.57$\\
                  & $\pm{0.51}$ & $\pm{0.6}$ & $\pm{0.58}$ & $\pm{0.45}$\\
\hline
Encoder+Prior+EKF & $-0.61$ & $-4.34$ & $-7.11$ & $-7.91$ \\
    & $\pm{0.21}$ & $\pm{0.28}$ & $\pm{0.35}$ & $\pm{0.3}$\\
\hline
\rowcolor{cyan}
\name & ${-2.4}$ & ${-5.45}$ & ${-7.91}$ & ${-8.54}$\\
\rowcolor{cyan}
      & $\pm{0.2}$ & $\pm{0.16}$ & $\pm{0.08}$ & $\pm{0.12}$\\
\hline
\end{tabular}}
\end{center}
\end{table}

%
%
\begin{table}[htbp!]
\label{Taylor Table}
\caption{Lorenz: Performance with partially domain knowledge. Mismatched Taylor expansion of state-evolution function, Fig \ref{fig:Miss-match_state_evolution_function.} }
\begin{center}
\scriptsize{
\begin{tabular}{|c|c|c|c|c|}
\hline
\rowcolor{lightgray}
Noise level $-\log(p_r)$ & 0.3 & 1 & 2 & 3 \\ [0.5ex]
\hline 
\hline
Encoder & $5.8$ & $-0.5$ & $-3.7$ & $-5.6$ \\
 & $\pm{1.5}$ & $\pm{0.9}$ & $\pm{1.1}$ & $\pm{1.3}$ \\
\hline
Encoder+Prior & $3.19$ & $-1.1$ & $-5.44$ & $-6.28$ \\
              & $\pm{0.83}$ & $\pm{0.8}$ & $\pm{0.73}$ & $\pm{0.7}$ \\
\hline
Encoder+Prior+EKF & $2.8$ & $-1.2$ & $-5.42$ & $-6.29$\\
                  & $\pm{0.59}$ & $\pm{0.7}$ & $\pm{0.61}$ & $\pm{0.56}$\\
\hline
RKN & $-1$ & $-4.2$ & $-6.9$ & $-7.8$\\
    & $\pm{2.3}$ & $\pm{2.1}$ & $\pm{1.5}$ & $\pm{1.1}$\\
\hline
\name~$J=2$ & ${-1.61}$ & ${-4.86}$ & ${-6.85}$  &  ${-7.53}$ \\
            & $\pm{0.3}$ & $\pm{0.29}$ & $\pm{0.42}$ & $\pm{0.31}$\\
\hline
\rowcolor{cyan}
\name~$J=5$ & ${-1.91}$ & ${-4.92}$ & ${-7.2}$ & ${-7.94}$\\
\rowcolor{cyan}
            & $\pm{0.38}$ & $\pm{0.2}$ & $\pm{0.41}$ & $\pm{0.3}$\\
\hline
\end{tabular}}
\end{center}
\end{table}

%
%
\begin{table}[htbp!]
\label{tbl:Decimation}
\caption{Lorenz: Performance with partially domain knowledge. Decimation: mismatch of sampling rate, Fig \ref{fig:Decimation.}}
\begin{center}
\scriptsize{
\begin{tabular}{|c|c|c|c|c|}
\hline
\rowcolor{lightgray}
Noise level $-\log(p_r)$ & 0.3 & 1 & 2 & 3 \\ [0.5ex]
\hline \hline
Encoder & $7.7$ & $1.78$ & $-1.95$ & $-3.11$ \\
        & $\pm{1.2}$ & $\pm{1.3}$ & $\pm{0.9}$ & $\pm{0.95}$ \\
\hline
Encoder+Prior & $6.1$ & $0.9$ & $-2.8$ & $-3.2$ \\
              & $\pm{0.72}$ & $\pm{0.7}$ & $\pm{0.53}$ & $\pm{0.6}$ \\
\hline
Encoder+Prior+EKF & $5.6$ & $0.8$ & $-2.83$ & $-3.21$\\
                  & $\pm{0.49}$ & $\pm{0.51}$ & $\pm{0.38}$ & $\pm{0.4}$\\
\hline
RKN & $4.3$ & $0.3$ & $-3.1$ & $-3.5$ \\
    & $\pm{1}$ & $\pm{1.3}$ & $\pm{1.5}$ & $\pm{2.1}$\\
\hline
\rowcolor{cyan}
\name & ${3.3}$ & ${-0.46}$ & ${-3.4}$ & ${-3.6}$\\
\rowcolor{cyan}
      & $\pm{0.09}$ & $\pm{0.21}$ & $\pm{0.18}$ & $\pm{0.16}$\\
\hline
\end{tabular}}
\end{center}
\end{table}


\end{document}